\documentclass[a4paper,11pt]{article}
\pdfoutput=1 

\usepackage{jcappub} 

\usepackage[T1]{fontenc}
\usepackage[utf8]{inputenc}
\usepackage{caption}
\usepackage{subcaption}

\usepackage{diagbox}
\usepackage{color}
\usepackage{multirow}
\usepackage{hhline}
\usepackage{tabularx}

\title{Space-borne atom interferometric gravitational wave detections. Part III. Eccentricity on dark sirens}

\author[a]{Tao Yang}
\emailAdd{yangtao.lighink@gmail.com}

\author[b,c,d]{Rong-Gen Cai}
\emailAdd{cairg@itp.ac.cn}

\author[a]{Hyung Mok Lee}
\emailAdd{hmlee@snu.ac.kr}

\affiliation[a]{Center for the Gravitational-Wave Universe, Astronomy Program Department of Physics and Astronomy, Seoul National University, 1 Gwanak-ro, Gwanak-gu, Seoul 08826, Korea}
\affiliation[b]{CAS Key Laboratory of Theoretical Physics, Institute of Theoretical Physics, Chinese Academy of Sciences, Beijing 100190, China}
\affiliation[c]{School of Physical Sciences, University of Chinese Academy of Sciences, No.19A Yuquan Road, Beijing 100049, China}
\affiliation[d]{School of Fundamental Physics and Mathematical Sciences, Hangzhou Institute for Advanced Study (HIAS), University of Chinese Academy of Sciences, Hangzhou 310024, China}

\abstract{
Eccentricity of the inspiraling compact binaries can greatly improve the distance inference and source localization of dark sirens. In this paper, we continue the research for the space-borne atom interferometric gravitational-wave detector AEDGE and investigate the effects of eccentricity on the dark sirens observed by AEDGE in the mid-band. We simulate five types of typical compact binaries with component mass ranging from $1-100~M_{\odot}$. The largest improvement for both distance inference and localization can be as much as 1.5--3 orders of magnitude. We then construct the catalogs of dark sirens observed by AEDGE in five years. We find eccentricity is crucial to the detection of golden binary black holes (BBH) whose host galaxy can be uniquely identified. With only 5--10 golden dark BBHs one can obtain a 2 percent precision measurement of $H_0$ which is sufficient to arbitrate the Hubble tension. Regardless of eccentricity, AEDGE can also observe tens of golden binary neutron stars (BNS) and neutron star--black hole binaries (NSBH) with unique host galaxies. These golden dark sirens can serve as early warnings for the follow-up observations of gravitational waves in the high frequency band as well as the search of their electromagnetic counterparts. Our results show eccentricity is a crucial factor in the detection, data analysis, and application of GWs with the atom interferometers in the mid-band.
}


\begin{document}
\maketitle
\flushbottom

\section{Introduction}
The ground-based gravitational wave (GW) detectors LIGO and Virgo have achieved great success in observing the GWs from the merger of binary neuron stars (BNS), binary black holes (BBH), and neutron star-black hole binaries (NSBH)~\cite{LIGOScientific:2016aoc,LIGOScientific:2018mvr,LIGOScientific:2020ibl,LIGOScientific:2021usb,LIGOScientific:2021djp}. The third-generation ground-based detectors such as ET~\footnote{\url{http://www.et-gw.eu/}} and CE~\footnote{\url{https://cosmicexplorer.org/}} are under design and construction. Together with the space-borne LISA~\footnote{\url{https://www.lisamission.org/}} and Chinese proposed projects like Taiji~\cite{Hu:2017mde,Ruan:2018tsw} and Tianqin~\cite{TianQin:2015yph} , they will be prepared for the detections of GWs around 2035. All of these are laser interferometers (LIs) and they will form a network of GW detectors from ground to space in the future. 

A novel type of GW detector called Atom interferometers (AIs) was proposed a decade ago~\cite{Dimopoulos:2007cj,Dimopoulos:2008sv,Graham:2012sy,Hogan:2015xla}. In the concept AIs, gravitational radiation is sensed through precise measurement of the light flight time between two distantly separated atomic inertial references, each in a satellite in Medium Earth orbit (MEO). Ensembles of ultra-cold atomic Sr atoms at each location serve as precise atomic clocks. Light flight time is measured by comparing the phase of laser beams propagating between the two satellites with the phase of lasers referenced to the Sr optical transitions~\cite{Graham:2017pmn}. Compared to the LIs, AIs consist of only a single baseline thus the design and building should be easier and cheaper than the traditional LIs. The AI projects such as ground based ZAIGA~\cite{Zhan:2019quq} in China, AION~\cite{Badurina:2019hst} in the UK, MIGA~\cite{Geiger:2015tma} in France, ELGAR~\cite{Canuel:2019abg} in Europe, and the space-borne MAGIS~\cite{Graham:2017pmn} and AEDGE~\cite{AEDGE:2019nxb} have been proposed and in preparation. 

AIs are proposed to probe not only the gravitational waves but also the dark matter. For GWs, AIs focus on the Deci-Hz gap between LIGO/Virgo and LISA.
In the mid-frequency range, one can observe the long inspiral period of BNS, BBH, and NSBH. During the long observation, the motion of the space-borne detector around the Sun as well as in Earth orbit would induce large Doppler and reorientation effects, providing a precise angular resolution. Based on the space-borne AEDGE, we compose a series of papers focusing on the GW detections by AIs. 
In the first paper~\cite{Cai:2021ooo} (hereafter Paper I) we forecast the bright sirens detected by AEDGE and their applications on cosmology. The specific analysis on the source localization for the dark sirens was conducted in the second paper~\cite{Yang:2021xox} (hereafter Paper II). The single baseline of AEDGE reorients on a rapid time scale compared to the observation duration. As a detector reorients and/or moves, the observed waveform and phase are modulated and Doppler-shifted. This allows efficient determination of sky position and polarization information~\cite{Graham:2017lmg,Graham:2017pmn}. In Paper II, we show AEDGE can even localize the dark sirens in such a small comoving volume that the unique host galaxy can be identified. These dark sirens are called ``golden dark sirens''. The measurements of Hubble constant from the simulated golden dark BNS and BBH are also performed. 

Many investigations suggest compact binaries that emit GWs can have non-negligible eccentricities and may contribute observational features in the sensitivity band of ground and space-based detectors~\cite{Antonini:2012ad,Samsing:2013kua,Thompson:2010dp,East:2012xq}. 
Different mechanisms of the dynamic formation of the compact binaries of black holes and neutron stars have been proposed to study their eccentricities~\cite{Rodriguez:2017pec,Samsing:2017xmd,Samsing:2017oij,Samsing:2018ykz,Wen:2002km,Pratten:2020fqn,OLeary:2008myb,Lee:2009ca}. 
Orbital eccentricity is arguably the most robust discriminator for distinguishing between isolated and dynamical BBH formation scenarios~\cite{Zevin:2021rtf}. Some studies indicate that a fraction of the binaries possess eccentricities larger than 0.1 at 10 Hz~\cite{Wen:2002km,Silsbee:2016djf,Antonini:2017ash,Liu:2019gdc}. The source localization improvement by the eccentricity for the ground-based detector networks has been investigated in some detail in~\cite{Sun:2015bva,Ma:2017bux,Pan:2019anf}. They found that the eccentricity has more distinct effects on localization for higher-mass binaries than for the lower ones. For the case of $100~M_\odot$ BBH, the improvement factor is about 2 in general when the eccentricity changes from 0.0 to 0.4~\cite{Pan:2019anf}. Such an improvement is not adequate to considerably shrink the uncertainty of host galaxies (redshift) of dark sirens in the LIGO/Virgo band, and to provide a conclusive measurement of the expansion of the Universe (e.g. the Hubble constant). While as shown in~\cite{Yang:2022tig}, the eccentricity can significantly improve the distance estimation and source localization in the mid-band. The multiple harmonics induced by eccentricity can break the degeneracy between parameters in the waveform. In addition, the higher modes can enter the detector band much earlier than the dominant mode, which can provide more angular information. At some specific orientations (inclination angles), the typical compact binaries can achieve $\mathcal{O}(10^2-10^4)$ improvement for the distance inference and $1.5\sim{3.5}$ orders of magnitude improvement for the sky localization. Such a huge improvement on the 3-D localization could dramatically shrink the uncertainty of the host galaxies of the dark sirens. Up to now, only GW190521 has been reported to be eccentric in the latest GW catalog GWTC-3~\cite{Romero-Shaw:2020thy,Gayathri:2020coq}. Considering the fact that the nonvanishing eccentricity is more likely to exist at lower frequency, we expect the dark sirens observed by the mid-band detector have greater potential on probing the cosmic expansion history, dynamics of dark energy, and gravity theory. 

In this paper, we extend our research in Paper II and take the eccentricity effects into account for the dark sirens with AEDGE. In Sec.~\ref{sec:typical}, we follow the methodology of~\cite{Yang:2022tig} to check the improvement of distance estimation and source localization by eccentricity for the typical BNS, NSBH, and BBH with AEDGE. In Sec.~\ref{sec:mock}, we adopt the similar method in Paper II to construct the catalogs of GWs for AEDGE. We first update the construction of catalogs of dark sirens in Paper II in the case of vanishing eccentricity. Comparing to Paper II, we update the waveform and the merger rates of BNS and BBH. We also include the NSBH into the simulation. We refine the fisher matrix calculation to ensure the convergence of the numerical derivatives. These updates and improvements make the simulation more realistic and reliable than that in Paper II. We then take account of eccentricity effects to construct the catalogs of dark sirens. We show how many potential host galaxies would be within the 3-D localization GW sources, with or without eccentricity. We estimate the the population of eccentric dark sirens and pick out the ones whose host galaxies can be best identified. We randomly select the golden dark sirens which AEDGE can track considering the limit of its operational time. The corresponding measurements of the Hubble constant are obtained in Sec.~\ref{sec:Hubble}. We give the conclusions and discussions in Sec.~\ref{sec:conclusion}.

\section{The improvement of distance estimation and localization from eccentricity \label{sec:typical}}
By adopting a similar strategy in~\cite{Yang:2022tig}, we mock up five types of typical compact binaries in GWTC-3~\cite{LIGOScientific:2021djp} with component mass ranging from $\mathcal{O}(1\sim100)~M_{\odot}$, i.e., a GW170817-like BNS with $(m_1,m_2)=(1.46,1.27)~M_{\odot}$, a GW200105-like NSBH with $(9.0,1.91)~M_{\odot}$, a GW191129-like light-mass BBH with $(10.7,6.7)~M_{\odot}$, a GW150914-like medium-mass BBH with $(35.6,30.6)~M_{\odot}$, and a GW190426-like heavy-mass BBH with $(106.9,76.6)~M_{\odot}$. Note the light, medium, and heavy mass are in the context of the stellar-mass binaries in GWTC-3. The redshifts (distances) are also consistent with the real events in the catalog. We sample 1000 random sets of the angular parameters from the uniform and isotropic distribution for each typical binary and assign six discrete initial eccentricities $e_0=0$, 0.01, 0.05, 0.1, 0.2, and 0.4 at $f_0=0.1$ Hz. Then we have $5\times 6\times 1000=3\times10^4$ cases. For each case, we perform the fisher matrix calculation to infer the errors of distance and sky location. 

We use {\sc PyCBC}~\cite{alex_nitz_2021_5347736} to generate the waveform with the non-spinning, inspiral-only EccentricFD waveform approximant available in {\sc LALSuite}~\cite{lalsuite}. EccentricFD corresponds to the enhanced post-circular (EPC) model in~\cite{Huerta:2014eca}. 
To the zeroth order in the eccentricity, the model recovers the TaylorF2 PN waveform at 3.5 PN order~\cite{Buonanno:2009zt}. To the zeroth PN order, the model recovers the PC expansion of~\cite{Yunes:2009yz}, including eccentricity corrections up to order $\mathcal{O}(e^8)$.
The strain can be written as~\cite{Huerta:2014eca}
\begin{equation}
\tilde{h}(f)=-\sqrt{\frac{5}{384}}\frac{\mathcal{M}_c^{5/6}}{\pi^{2/3}d_L}f^{-7/6}\sum_{\ell=1}^{10}\xi_{\ell}\left(\frac{\ell}{2}\right)^{2/3}e^{-i\Psi_{\ell}} \,.
\label{eq:epc}
\end{equation} 
The waveform keeps up to 10 harmonics, which corresponds to a consistent expansion in the eccentricity to $\mathcal{O}(e^8)$ both in the amplitude and in the phase~\cite{Yunes:2009yz}. In the vanishing eccentricity case, only the dominant (quadrupole) mode $\ell=2$ remains, which is identical to the circular TaylorF2 model. With nonvanishing eccentricities, the induced multiple harmonics make the distance and angular parameters nontrivially coupled, enabling us to break the degeneracy among these parameters. In addition, the frequency of each harmonics is $\ell F$ with $F$ the orbital frequency. Thus the higher harmonics ($\ell>2$) should enter the detector band much earlier than the dominant mode ($\ell=2$), which can provide more angular information. The $\xi_{\ell}$'s depend on the antenna pattern functions (also called detector response functions) $F_{+,\times}$. For the space-borne AEDGE, we should consider the motion of the detector thus $F_{+,\times}$ are functions of time. We give the detailed calculation of the antenna pattern functions in appendix~\ref{app:F}. 

We have 11 parameters in the waveform, namely the chirp mass $\mathcal{M}_c$, the symmetric mass ratio $\eta$, the luminosity distance $d_L$, the inclination angle $\iota$, the sky location ($\theta$, $\phi$), the polarization $\psi$, the time and phase at coalescence ($t_c$, $\phi_c$), the initial eccentricity $e_0$ at frequency $f_0$, the azimuthal component of inclination angles (longitude of ascending nodes axis) $\beta$. To estimate the uncertainty and covariance of the waveform parameters, we adopt the Fisher matrix technique 
\begin{equation}
\Gamma_{ij}=\left(\frac{\partial h}{\partial P_i},\frac{\partial h}{\partial P_j}\right)\,,
\end{equation}
with $P_i$ one of the 11 waveform parameters.
The inner product is defined as
\begin{equation}
(a,b)=4\int_{f_{\rm min}}^{f_{\rm max}}\frac{\tilde{a}^*(f)\tilde{b}(f)+\tilde{b}^*(f)\tilde{a}(f)}{2 S_n(f)}df\,.
\label{eq:innerp}
\end{equation}
For the noise power spectral density (PSD) $S_n(f)$, we adopt the sensitivity curve of AEDGE in the resonant modes (see the envelope in figure 1 of~\cite{Ellis:2020lxl}).
Then the covariance matrix of the parameters is $C_{ij}=(\Gamma^{-1})_{ij}$, from which the uncertainty of each parameter $\Delta P_i=\sqrt{C_{ii}}$. The error of the sky localization is~\cite{Cutler:1997ta}
\begin{equation}
\Delta \Omega=2\pi |\sin(\theta)|\sqrt{C_{\theta\theta}C_{\phi\phi}-C_{\theta\phi}^2}\,.
\end{equation}
We calculate the partial derivatives $\partial \tilde{h}/\partial P_i$ numerically by $[\tilde{h}(f,P_i+dP_i)-\tilde{h}(f,P_i)]/dP_i$, with $dP_i=10^{-n}$.  For each parameter, we need to optimize $n$ to make the derivative converged so that the Fisher matrix calculation is reliable. 

For each typical event, the chirp mass $\mathcal{M}_c$, symmetric mass ratio $\eta$, and distance $d_L$ are calculated from the component mass and redshift. The angular parameters $P_{\rm ang}=\{\iota,~\theta,~\phi,~\psi,~\beta\}$ are sampled from the uniform and isotropic distribution with 1000 sets for each typical event. We use the inclination angle $\iota$ to represent to angular parameter since we find it is more relevant in terms of the results. Without loss of generality, we fix the coalescence time and phase to be $t_c=\phi_c=0$. We choose the frequency band of AEDGE to be [0.1, 3] Hz, where the detector is the most sensitive. This range corresponds to lower and upper bounds of frequency in the integral of Eq.~(\ref{eq:innerp})~\footnote{This is also different with Paper II in which we naively set the lower bound of frequency to be 0.2 for BNS and 0.05 for BBH.}. However, we should consider the limited operation time of AEDGE for tracking the GWs. We set quadrupole ($\ell=2$) as the reference mode and its frequency is double of the orbital's, $f_{\ell=2}=2F$. Then the evolution of the binary orbit can be calculated in terms of the quadrupole frequency. To ensure the observational time of AEDGE for all harmonics is around 400 days ($\sim1$ year), we set the starting frequency of quadrupole $f_{\rm start}(\ell=2)$ to be 0.2, 0.1, 0.059, 0.026, and 0.0105 Hz for the typical BNS, NSBH, light BBH, medium BBH, and heavy BBH, respectively. That is, for the strain Eq.~(\ref{eq:epc}) we should neglect the contribution of all the harmonics when the quadrupole's frequency is smaller than $f_{\rm start}(\ell=2)$~\footnote{Note in the mid band we can only observe the inspiral phase of these binaries, thus we do not need to care about the upper frequency limit at the innermost-stable circular orbit.}. Thus we multiply the strain by the step function
\begin{equation}
\tilde{h}_{\rm AEDGE}(f)=\tilde{h}(f)\mathcal{H}(2f-\ell f_{\rm start}) \,,
\label{eq:hAEDGE}
\end{equation}
with the unit step function
\begin{equation}
\mathcal{H}(x)=
\begin{cases}
1 & {\rm if}~x\geq0 \,, \\
0 & {\rm otherwise} \,.
\end{cases}
\end{equation}
For the orbital phase evolution, we numerically solve Eqs. (3.11) and (4.24) in~\cite{Yunes:2009yz} to obtain the time to coalescence $t(f)$ for a nonvanishing $e_0$. The time to coalescence at a specific frequency is smaller for a larger eccentricity. So for the fixed $f_{\rm start}(\ell=2)$, the observational time is shorter with a larger eccentricity.

We collect all the results of the fisher matrix for the $3\times10^4$ cases. Same as~\cite{Yang:2022tig}, for each typical event with a specific orientation, we define the ratios
\begin{equation}
R_{\Delta d_L}=\frac{\Delta d_L|_{e_0={\rm nonzero}}}{\Delta d_L|_{e_0=0}}~{\rm and}~R_{\Delta \Omega}=\frac{\Delta \Omega |_{e_0={\rm nonzero}}}{\Delta\Omega|_{e_0=0}} \,,
\end{equation} 
to show the improvement induced by eccentricity in that orientation. If $R<1$, there is an improvement in the relevant parameter. A smaller $R$ indicates a  larger improvement. We show the scatter plots of  $\Delta d_L/d_L$, $R_{\Delta d_L}$, $\Delta \Omega$, and $R_{\Delta \Omega}$ against $\iota$. To give the statistical results, we define the minimum, mean, and maximum value of $x$ in the 1000 orientations as $\min(x)$, $\mathbb{E}(x)$, and $\max(x)$, respectively.

\begin{figure}
\centering
\includegraphics[width=0.49\textwidth]{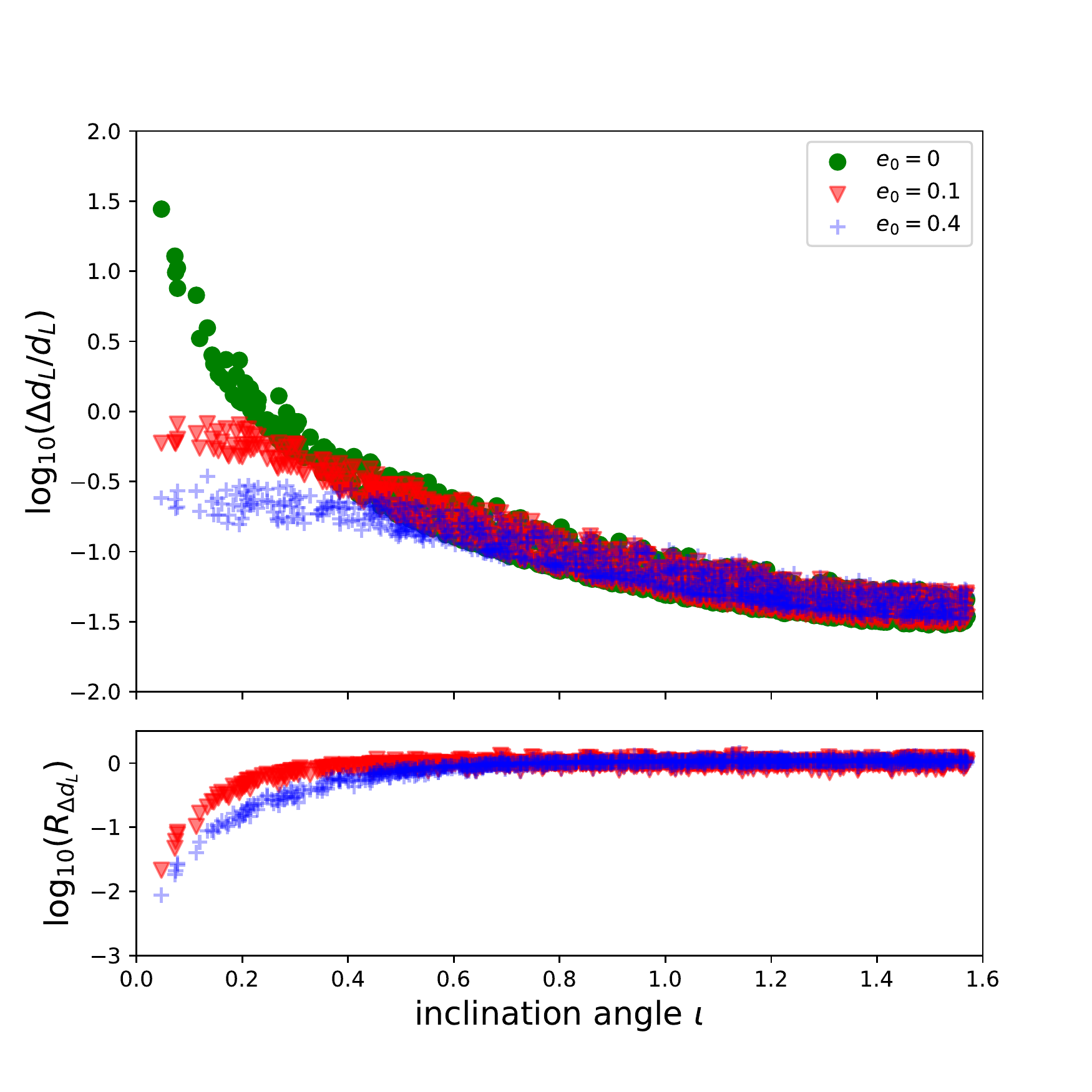}
\includegraphics[width=0.49\textwidth]{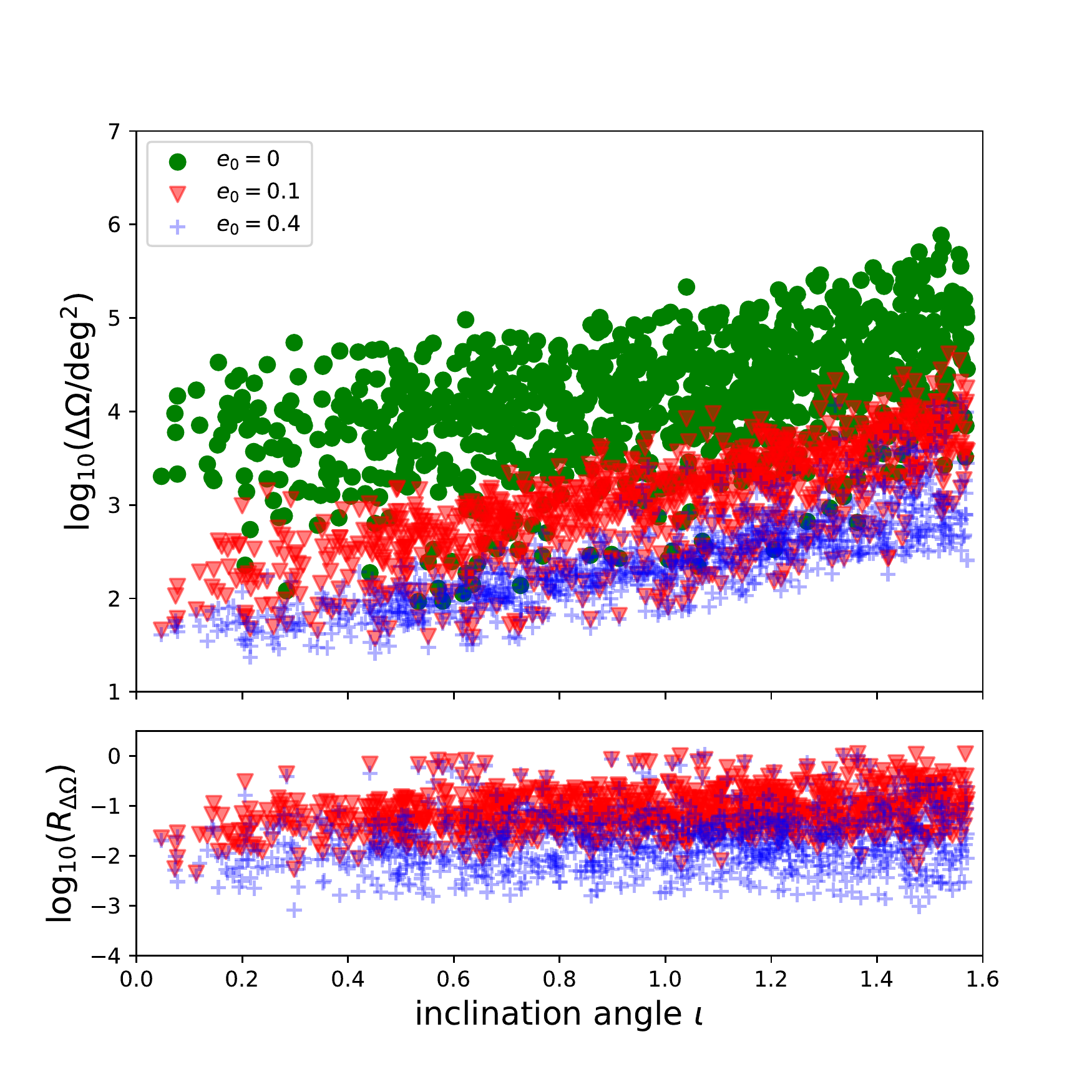} 
\caption{The distance inference of GW170817-like BNS (left) and source localization of GW190426-like heavy BBH.}
\label{fig:rep}
\end{figure}

In figure~\ref{fig:rep}, we only show the distance inference of GW170817-like BNS and source localization of GW190426-like heavy BBH to represent
our main results. We just compare the cases with $e_0=$0, 0.1, and 0.4 to give a concise look.  The complete results can be found in  appendix~\ref{app:sup}. As shown in left panel of figure~\ref{fig:rep}, a nonvanishing eccentricity can significantly improve the distance inference in the near face-on orientations (small inclination angle). Among all 1000 orientations, the $\max(\Delta d_L/d_L)$ of GW170817-like BNS is reduced from 27.74 ($e_0=0$) to $0.82$ ($e_0=0.1$) and $0.35$ ($e_0=0.4$). Comparing to $e_0=0$ case, the largest improvement ($\min(R_{\Delta d_L})$) corresponds to 47 and 115 times stricter with $e_0=0.1$ and $e_0=0.4$, respectively. The huge improvement of distance inference in the near face-on orientations is true for all the typical events. The binaries with larger component mass and eccentricity can achieve more improvement. As shown in appendix~\ref{app:sup}, for the heavy BBH with $e_0=0.4$, $\min(R_{\Delta d_L})=0.0012$, corresponding to 833 times improvement. We also find that for the heavy BBH case, there is an overall improvement of distance inference in all orientations. Our results indicate that the eccentricity effects are more distinct for the larger mass compact binaries. 

For the source localization, we find eccentricity can lead to significant improvement for the BBH cases which have larger component mass than BNS and NSBH cases. As shown in the right panel of figure~\ref{fig:rep}, the localization of heavy BBH is significantly improved by the eccentricities in almost all orientations. The largest improvement $\min(R_{\Delta \Omega})= 8.16\times 10^{-4}$, corresponding to $1.23\times10^3$ times tighter. Like the distance inference, the heavier binaries benefits more from the eccentricity for the source localization. 
The details of the improvement by eccentricity can also be found in the figures summarized in appendix~\ref{app:sup}.

\begin{figure}
\centering
\includegraphics[width=0.7\textwidth]{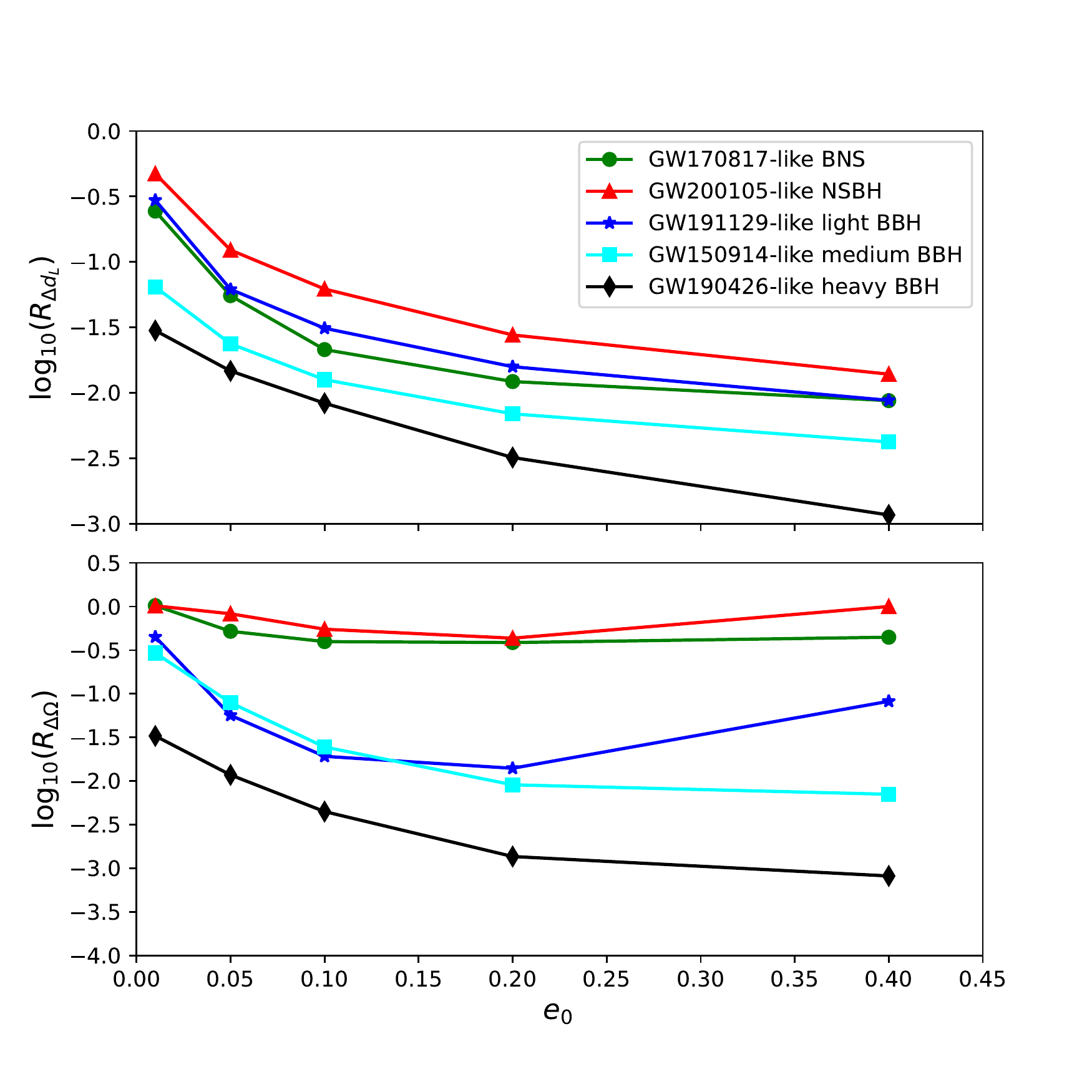}
\caption{The largest improvement of distance inference (upper panel) and source localization (lower panel) for the typical compact binaries in 1000 orientations with variable eccentricities.}
\label{fig:Rwe}
\end{figure}

To illustrate the improvement of distance inference and localization for these typical binaries with variable eccentricities, we show the largest improvement ($\min(R)$ in 1000 orientations) of each case in Fig.~\ref{fig:Rwe}.  We can see generally a heavier binary with higher eccentricity can achieve more improvement of distance inference and source localization. With eccentricity $e_0=0.4$, these typical binaries can most achieve 
1.5--3 orders of magnitude
improvement for the distance inference (from BNS to heavy BBH). As for the source localization, BNS and NSBH can not benefit much from the eccentricity. While BBHs can most achieve 1.5--3 orders of magnitude improvement (from light BBH to heavy BBH). We should note some anomalies in figure~\ref{fig:Rwe}. 1) For the distance inference, the typical BNS benefits more from eccentricities than the typical NSBH and light BBH do. 2) For the source localization, BNS behaves similarly with NSBH and both have almost no improvement from eccentricity. 3) The light BBH's tendency is very close to that of medium BBH when $e_0<0.2$ and then they diverge for larger eccentricity. 4) In the BNS, NSBH, and especially for the light BBH cases, the localization achieves largest improvement when $e_0=0.2$, a higher eccentricity ($e_0=0.4$) can even worsen the performance. These anomalies are caused by many factors. On the one hand, eccentricity adds more harmonics in GWs. These harmonics can enlarge the SNR and improve the parameter estimation. The higher modes which enter the detector band much earlier can provide more angular information. On the other hand, eccentricity shrinks the inspiral time within the frequency band, which could lower the SNR and hence worsen the parameter estimation and localization. In addition, due to the different starting frequencies, for each binaries the detector band cover different length of harmonics. For instance, in BNS case, at the starting frequency $f_{\rm start}(\ell=2)=0.2$, all harmonics including $\ell=1$ falls inside the detector band (0.1--3Hz). But in NSBH case, at $f_{\rm start}(\ell=2)=0.1$, the $\ell=1$ mode's frequency is 0.05, falling outside the detector band hence should be truncated. Moreover, we have two more parameters $e_0$ and $\beta$ in the eccentric waveform, which could degrade the overall performance of the parameter estimation. All above factors compete with each other and make the parameter estimation (distance inference and localization) differ from case to case. 

Here we would like to provide some more explanations for the inverse tendency of the error of the distance and localization versus the inclination angle. We can see in figure~\ref{fig:rep} that generally the the error of distance is larger in smaller orbital inclination. On the contrary, the source localization is better when inclination is smaller. For distance it is due to the degeneracy between distance and inclination angle. In the amplitude of GW waveform~$h\sim \mathcal{A}_++\mathcal{A}_\times$, the distance $d_L$ and inclination angle $\iota$ are tangled in the plus and cross polarization with different form, $\mathcal{A}_+\sim\frac{1}{d_L}\frac{1+\cos(\iota)}{2}$ and $\mathcal{A}_+\sim\frac{1}{d_L}\cos(\iota)$. In order to identify the inclination of the binary system using the polarizations of the gravitational wave, we must distinguish the contributions of the plus and cross polarizations. At small $\iota$, the two amplitudes from plus and cross polarizations have nearly identical contributions to the overall gravitational-wave amplitude. This is the main factor that leads to the strong degeneracy in the measurement of the distance and inclination~\cite{Usman:2018imj}. So we expect a larger degeneracy between $d_L$ and $\iota$ in the near face-on orientations and hence larger errors for both distance and inclination angle. As for the source localization, there is no obvious degeneracy between the sky location parameters $(\theta,\phi)$ and inclination angle $\iota$. However, at smaller $\iota$ the SNR is larger. So the parameter estimation should be better than that at larger $\iota$. 

We have showed that eccentricity, which is more likely to exist in the mid-band that in LIGO/Virgo band, can improve the distance inference and source localization of dark sirens with AEDGE significantly. Note GWs are best localized at smallest orbital inclination where the distance are worst determined. But eccentricity happens to improve the distance inference most significantly there. In addition, one of the main targets for the mid-band detector like AEDGE is the intermediate mass black holes (IMBH). While in this paper we showed that the heaviest BBH can benefit most from the eccentricity for the distance inference and source localization. These facts suggests that eccentricity is the perfect ingredient for AEDGE dark sirens as precise probes of the Universe 

\section{The construction of dark sirens catalogs and the host galaxy identification \label{sec:mock}}
Considering the fact that eccentricity plays an important role in the distance inference and source localization of dark sirens with the mid-band detector AEDGE, we should take eccentricity effects into the construction of the dark sirens catalogs. In this section, we first update the construction of the catalogs of dark sirens in Paper II which does not consider the eccentricity effects, i.e., $e_0=0$. We adopt the EccentricFD waveform in which the $e_0=0$ case is equivalent to TaylorF2 at 3.5 PN order. While, in paper II we only expand the waveform to 2 PN order in the phase. We also update the BNS and BBH merger rates from the latest GWTC-3, as well as the BBH population. In addition to BNS and BBH, we add NSBH catalog. More importantly, we refine the numerical derivatives for the fisher matrix calculation, which would make the results more stable and thus robust and reliable. Then we include the eccentricity effects into the construction of the catalogs, to assess the its influence on the population and localization of the binaries.

We follow Paper II and assume the formation of compact binaries tracks the star formation rate.
The merge rate per comoving volume at a specific redshift $R_m(z_m)$ is related to the formation rate of massive binaries and the time delay distribution $P(t_d,\tau)=\frac{1}{\tau}\exp(-t_d/\tau)$ with an e-fold time of $\tau=100$ Myr~\cite{Vitale:2018yhm},
\begin{equation}
R_m(z_m)=\int_{z_m}^{\infty}dz_f\frac{dt_f}{dz_f}R_f(z_f)P(t_d) \,.
\label{eq:Rm}
\end{equation}
Here $t_m$ (or the corresponding redshift $z_m$) and $t_f$ are the look-back time when the systems merged and formed. $t_d=t_f-t_m$ is the time delay. $R_f$ is the formation rate of massive binaries and we assume it is proportional to the Madau-Dickinson (MD) star formation rate~\cite{Madau:2014bja},
\begin{equation}
\psi_{\rm MD}=\psi_0\frac{(1+z)^{\alpha}}{1+[(1+z)/C]^{\beta}} \,,
\label{eq:psiMD}
\end{equation}
with parameters $\alpha=2.7$, $\beta=5.6$ and $C=2.9$. The normalization factor $\psi_0$ is determined by the local merger rates. We adopt the local merger rates of BNS, NSBH, and BBH inferred from GWTC-3, with $\mathcal{R}_{\rm BNS}=105.5^{+190.2}_{-83.9}~\rm Gpc^{-3}~\rm yr^{-1}$, $\mathcal{R}_{\rm NSBH}=45^{+75}_{-33}~\rm Gpc^{-3}~\rm yr^{-1}$, and $\mathcal{R}_{\rm BBH}=23.9^{+14.3}_{-8.6}~\rm Gpc^{-3}~\rm yr^{-1}$~\cite{LIGOScientific:2021psn}. Note we assume the observed NSBH GW200105 and GW200115 are representatives of the population of NSBH. Then we convert the merger rate per comoving volume in the source frame to merger rate density per unit redshift in the observer frame
\begin{equation}
R_z(z)=\frac{R_m(z)}{1+z}\frac{dV(z)}{dz} \,,
\label{eq:Rz}
\end{equation}
where $dV/dz$ is the comoving volume element. 

Having the merger rates as redshift, we can sample the redshift distribution of BNS, NSBH, and BBH. Like Paper II, we use the median merger rates to construct the catalogs. We have 11 parameters in the waveform (for vanishing eccentricity there are 9 except $e_0$ and $\beta$). The luminosity distance $d_L$ is calculated from the sampled redshift by assuming a fiducial cosmological model $\Lambda$CDM with $H_0=67.72~\rm km~s^{-1}~Mpc^{-1}$ and $\Omega_m=0.3104$, corresponding to the mean values obtained from the latest \textit{Planck} experiment~\cite{Planck:2018vyg}. The sky localization ($\theta$, $\phi$), inclination angle $\iota$, and polarization $\psi$ are drawn from isotropic distribution. Without loss of generality we set the time and phase at coalescence to be $t_c=\phi_c=0$. As for the chirp mass and symmetric mass ration, we consider different strategy for these three binary types. In the BNS case, we assume a uniform distribution of mass in [1, 2.5] $M_{\odot}$, which is consistent with the assumption for the prediction of the BNS merger rate in GWTC-3~\cite{LIGOScientific:2021psn}. In the NSBH case, since the merger rate is inferred by assuming the observed NSBH GW200105 and GW200115 are representatives of the population of NSBH, we just randomly choose the component mass of these two events. As for the BBH case, we adopt the same strategy in Paper II with BBH population in GWTC-3. We draw the distribution of component mass of BBH from the histogram of mass distribution of BBH in GWTC-3~\footnote{We first infer the histograms of primary mass $m_1$ and mass ratio  $q$ from GWTC-3. The distribution of $m_1$ and $q$ are sampled accordingly. Then the second mass is just $m_2=m_1q$. We should make sure that $m_2\ge3~M_{\odot}$.}. The primary mass and mass ratio peak around 30--40 $M_{\odot}$ and 0.7. 

We sample the mergers of BNS, NSBH, and BBH in 5 years since the operation time of AEDGE is supposed to be 5--10 years~\cite{AEDGE:2019nxb}. We set the frequency band and starting frequency to be same as in section~\ref{sec:typical}. This means the observational time for each event is around 1 year. For each sampled merger, we assume four discrete eccentricities, i.e., $e_0=0$, 01, 0.2, and 0.4 at $f_0=0.1$ Hz. We select the mergers with SNR>8 as the candidate events that could be detected (within the detection range) by AEDGE in 5 years. For each events, we adopt the fisher matrix to derive their distance errors and source localizations. By assigning a uniform eccentricity for each event, we would like to assess the influence of eccentricity on the population and localization of the GWs that could be detected by AEDGE. We will give a discussion about the distribution of eccentricity and the realistic population of eccentric binaries later.

\begin{figure}
\centering
\includegraphics[width=\textwidth]{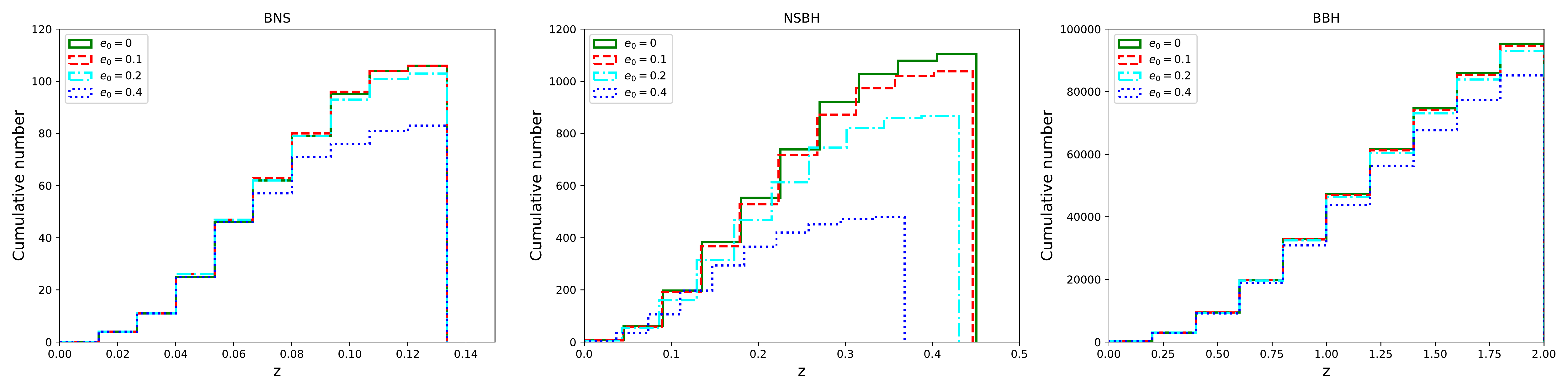}
\caption{The cumulative histogram of events which are within the detection range of AEDGE in 5 years. Note we set a cut-off at $z=2$ for BBH.}
\label{fig:hist}
\end{figure}

Figure~\ref{fig:hist} shows the cumulative histogram of events within the detection range of AEDGE in 5 years. The highest redshift AEDGE can reach for BNS and NSBH are around 0.13 and 0.45, respectively. For BBH, the horizon is much larger but we set a cut-off at $z=2$ since the for higher redshift we usually can not obtain the spectroscopic measurement of the redshift. In addition, the large uncertainty of localization makes the BBH at high redshift useless for our purpose in this paper. In the circular case, the total numbers are 106, 1105, and 95369 for BNS, NSBH, and BBH ($z\leq2$), respectively. The numbers of BNS and BBH are smaller than that in Paper II, which is due to the different choice of the merger rates and the lower limit of frequency band of AEDGE (we adopt $f_{\rm min}=0.05$ Hz for BBH in Paper II, while in this paper $f_{\rm min}=0.1$ Hz). We note that a larger eccentricity leads a smaller population of the events. This is due to the fact that eccentricity reduces the inspiral (orbital evolution) time of binaries in the frequency band (0.1--3 Hz). The smaller observational time leads smaller accumulation of SNR, especially for the dominant quadrupole mode. So the GWs whose SNR are just a little above the detection threshold when $e_0=0$ may not be detected if they have nonvanshing eccentricities. In the NSBH case, the largest redshift AEDGE can reach is smaller for eccentric events. Comparing to BNS and BBH, the population of NSBH decrease the most with eccentricity. The reason is that we choose GW200105 and GW200115 as the representatives of NSBH population. The component masses in the NSBH catalog are fixed to be the same as either of these two typical events. So, the high-redshift eccentric events are definitely to be below the SNR threshold. While for BNS and BBH, there may be a larger sampled component mass to compensate for the low SNR. 

\begin{figure}
\centering
\includegraphics[width=\textwidth]{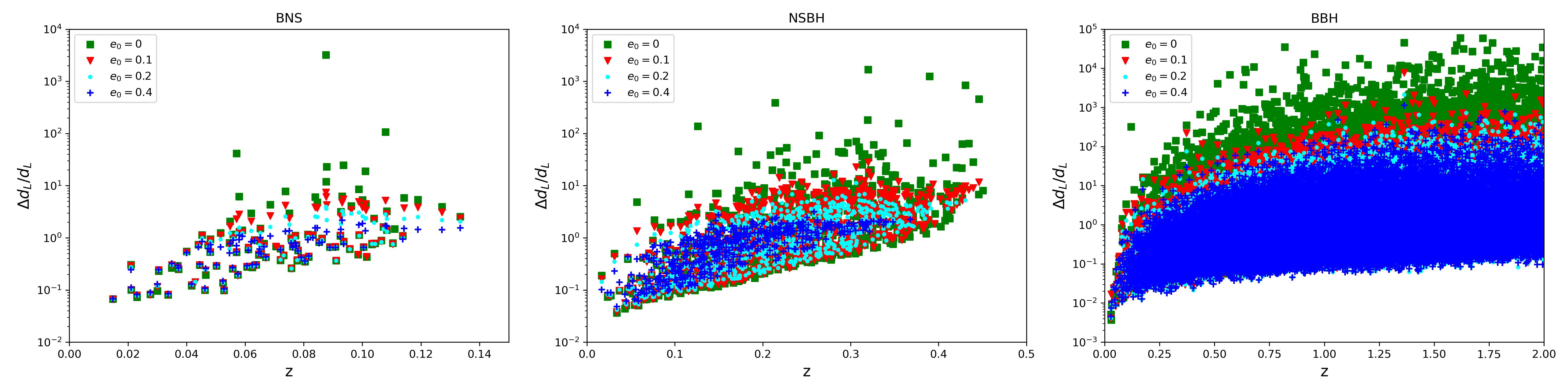}
\caption{The distance error of the events which are within the detection range of AEDGE in 5 years.}
\label{fig:err_dL}
\end{figure}

\begin{figure}
\centering
\includegraphics[width=\textwidth]{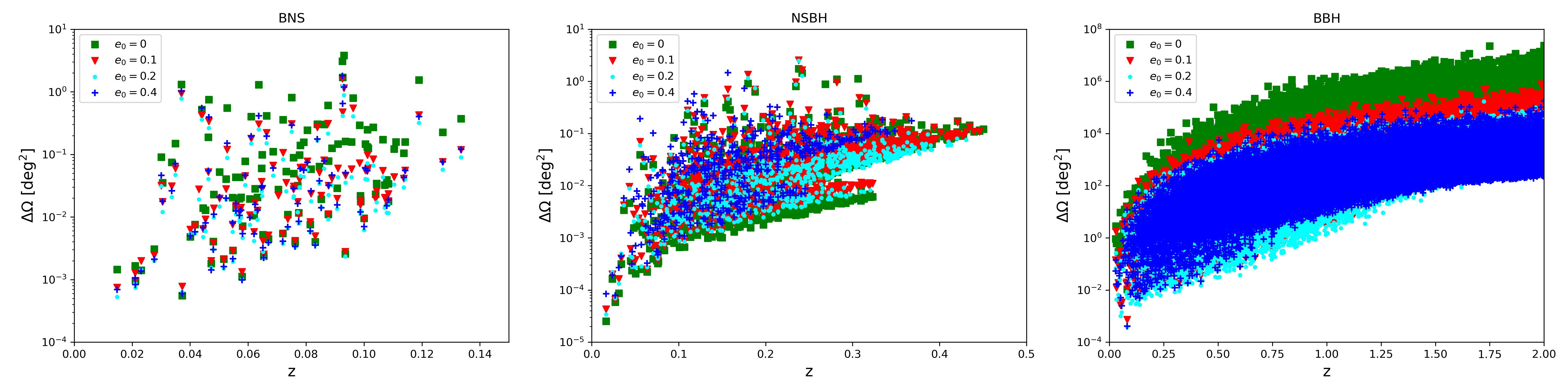}
\caption{The localization of the events which are within the detection range of AEDGE in 5 years.}
\label{fig:err_Omega}
\end{figure}

Figures~\ref{fig:err_dL} and~\ref{fig:err_Omega} show the error of distance and localization of the binaries that are within the detection range of AEDGE in 5 years. Eccentricity can significantly improve the overall distance inference of the binaries in the catalogs. For the source localization, BNS and NSBH can not benefit obviously from eccentricity. The localizations of eccentric events are even worsen in some cases. However, the source localization of BNS and NSBH are $\mathcal{O}(10^{-4})~\rm deg^{2}$ level even without eccentricity. While BBH's localization is considerably improved by eccentricity. The optimal localization at low redshift is improved to be better than $\mathcal{O}(10^{-3})~\rm deg^{2}$.  We find that, in some cases, with $e_0=0.2$ the binaries can achieve the most improvements. All of these features can be expected based on the results in section~\ref{sec:typical}.

To assess the galaxy identification of the binaries in the catalogs, we should calculate their 3-D localization volumes which can be obtained from the errors of distance and localization in figures~\ref{fig:err_dL} and~\ref{fig:err_Omega}. We follow the method in~\cite{Yu:2020vyy} to convert $\Delta d_L$ and $\Delta\Omega$ to the 99\% confidence ellipsoid of the localization. We use $V_{\rm loc}$ to denote the 3-D volume of the localization. To estimate the numbers of potential host galaxies in the localization volume, we assume the galaxy is uniformly distributed in the comoving volume and the number density $n_g= 0.01~\rm Mpc^{-3}$. This number is derived by taking the Schechter function parameters in B-band $\phi_*=1.6\times 10^{-2} h^3 {\rm Mpc^{-3}}, \alpha=-1.07, L_*=1.2\times 10^{10} h^{-2} L_{B,\odot}$ and $h=0.7$, integrating down to 0.12 $L_*$ and comprising 86\% of the total luminosity~\cite{Chen:2016tys}. Then the threshold localization volume is $V_{\rm th}=100~\rm Mpc^3$. If $V_{\rm loc}\leq V_{\rm th}$, the host galaxy of the dark sirens can be identified uniquely and we call these golden dark sirens.

\begin{figure}
\centering
\includegraphics[width=\textwidth]{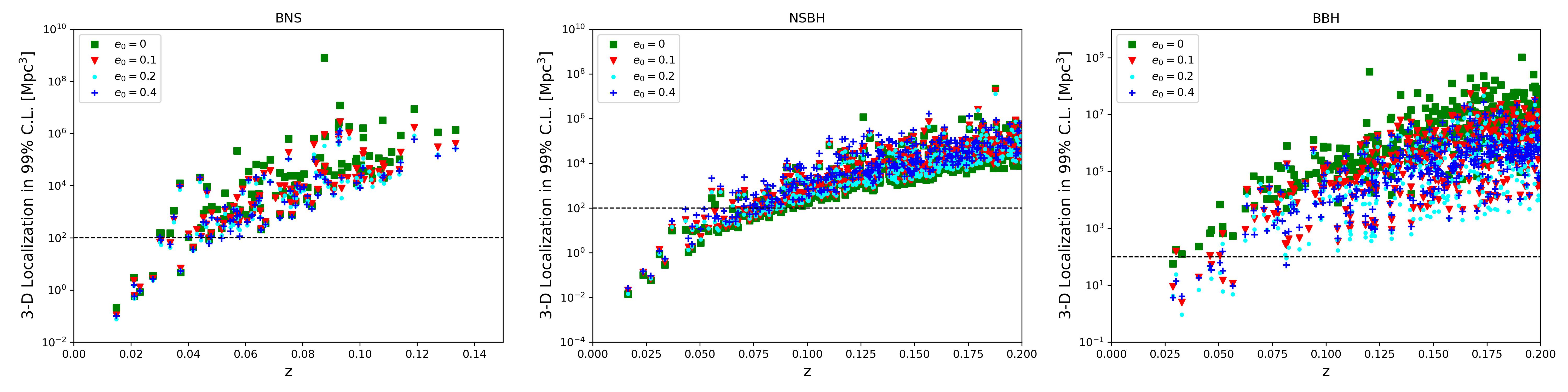}
\caption{The 3-D localization volumes of the events which are within the detection range of AEDGE in 5 years. The horizontal dashed line corresponds to the threshold volume that the unique host galaxy can be identified.}
\label{fig:V_loc}
\end{figure}

Figure~\ref{fig:V_loc} shows the the 99\% confidence level (C.L.) of the 3-D localization of the events that are within the detection range of AEDGE in 5 years. We can see in the circular case, several BNS and NSBH events at low redshift can be localized within $V_{\rm th}$. As for BBH, a few events can be localized with only one potential host galaxy. However, through the improvement from eccentricity, the eccentric BBH at low redshift can be well localized to become the golden dark sirens. The number of the golden dark sirens in the catalogs are summarized in table~\ref{tab:np}. We also show the number of dark sirens whose potential host galaxies' count $n_p$ are less than 10. The result shows BBH can benefit the most from the eccentricity. In the circular case, it is almost impossible to detect the golden dark BBH. While nonvanishing eccentricities significantly increase the possibility to detect the golden BBH at low redshift. Note in the NSBH case, eccentricity would worsen the results compared to the circular case. The $e_0=0.2$ case gives an overall better result than other cases. All of these results are consistent with the expectation in section~\ref{sec:typical}.

\begin{table}
\centering
\resizebox{\columnwidth}{!}{
\begin{tabular}{|c|c|c|c|c|c|c|c|c|} 
\hline
            & \multicolumn{2}{c|}{$e_0=0$} & \multicolumn{2}{c|}{$e_0=0.1$} & \multicolumn{2}{c|}{$e_0=0.2$} & \multicolumn{2}{c|}{$e_0=0.4$}  \\ 
\hline
Binary type & Golden~ & $1<n_p\leq 10$     & Golden & $1<n_p\leq 10$        & Golden & $1<n_p\leq 10$        & Golden & $1<n_p\leq 10$         \\
BNS         & 8       & 17                 & 11     & 23                    & 12     & 26                    & 11     & 25                     \\
NSBH        & 46      & 69                 & 38     & 61                    & 40     & 61                    & 29     & 48                     \\
BBH         & 1       & 7                  & 6      & 14                    & 10     & 22                    & 10     & 9                      \\ 
\hline
Total       & 55      & 93                 & 55     & 98                    & 62     & 109                   & 50     & 82                     \\
\hline
\end{tabular}
}
\caption{The number of golden dark sirens which are within the detection range of AEDGE in 5 years. We also list the number of dark sirens whose potential host galaxies' count are less than 10. We assume the number density of galaxies $n_g= 0.01~\rm Mpc^{-3}$.}
\label{tab:np}
\end{table}

As shown in table~\ref{tab:np}, the numbers of golden dark BNS and BBH in $e_0=0$ case are smaller than those of Paper II. The difference in the numbers arises from several factors. (1) In this paper, we adopt an updated median merger rates from GWTC-3 which is lower than that in Paper II from GWTC-2. (2) We use the EPC waveform which is identical to TaylorF2 with 3.5 PN orders when eccentricity is 0. While in paper II we use the waveform with only 2 PN orders. (3) We set the lower limit of frequency of AEDGE to be 0.1 Hz for BBH. Thus the inspiral time (thus the accumulated SNR) of quadrupole mode is smaller compared to Paper II in which we set $f_{\rm low}=0.05$ Hz. (4) We refine the Fisher matrix calculation to make the numerical derivatives more stable and reliable. All of these factors make the numbers in this paper more realistic and conservative compared to the optimistic estimation in Paper II. However, the application of AEDGE only depends on a few (5--10) golden dark sirens that it can track during the mission time. In this paper, we find that AEDGE can still observe 50--60 dark sirens regardless of the eccentricity. So the difference in the numbers of golden dark BNS and BBH will not influence the main result for the application of AEDGE on cosmology. Later we can find the constraints of the Hubble constant from 5--10 dark sirens in this paper are consistent with that in Paper II.

In this section, we construct the catalogs of GWs that could be detected by AEDGE in 5 years. However, as discussed in Paper II, in the resonant mode AEDGE can only track one event at a time. So only a small fraction of the total events in the catalogs can be observed by AEDGE in 5 years. Thus we should take the priority of observation into account. In this paper, we focus on the dark sirens with the best localization. Thus gold dark sirens are the sources we would like to track first with AEDGE. Since we set the observational time for one event is around 1 year, the total number of dark sirens AEDGE can track during its mission time (5--10 years) is 5--10. As shown in figure~\ref{fig:V_loc}, the possibility of observing golden dark sirens at $z\leq0.05$ is pretty high. The total number of golden dark sirens in table~\ref{tab:np} is from 50--60, for either circular or eccentric cases. We can rely on the real-time data analysis in the tracking process to fast identify the golden dark sirens. We need to predict the quality (also the properties) of the event as soon as possible such that we can improve the successful capture of the golden events.

When constructing the catalogs of eccentric GWs, we assume a uniform (average) eccentricity for all the events in the catalogs. By doing so we just would like to assess the influence of eccentricity on the population of GWs (and golden dark sirens). However, this is not a realistic assumption. The formation channel of compact binaries and distribution of eccentricity are still under debate and not clear~\cite{Wen:2002km,Kowalska:2010qg,Takatsy:2018euo}.  Simulations suggest that 10\% of the BBH population is formed dynamically and at least half of them have eccentricity larger than 0.1 at 10 Hz~\cite{Samsing:2013kua,Samsing:2017xmd,Samsing:2017rat} (and references therein). As a rough estimation, most of these dynamical BBH should hold considerably high eccentricity at 0.1 Hz. For the isolated formation scenarios, even binaries born with high eccentricity could be fully circularized when entering the LIGO/Virgo band, by efficiently damping orbital eccentricity through angular momentum loss from GW emission. However, the eccentricity of the field binary at mid-band is still uncertain. Nevertheless, the probability of a nonvanishing eccentricity should be much higher at mid-band than at LIGO/Virgo band. To summary, we believe at least more than 10\% population of the binaries should be eccentric at mid-band. To detect the golden dark BBH with AEDGE, a nonvanishing eccentricity is crucial. From table~\ref{tab:np}, the total number of golden dark sirens are around 50--60 for either vanishing or nonvanishing eccentricity. Therefore, tracking 5--10 golden dark sirens with AEDGE in 5 years should be ensured regardless of the distribution of eccentricity. 

\section{The Hubble constant from golden dark sirens \label{sec:Hubble}}

In this section we would like to estimate the constraint ability on cosmological parameters from the the eccentric dark sirens detected by AEDGE in 5--10 years. Since the golden dark sirens observed by AEDGE only reside at low redshift ($z<1$), we only forecast their applications on the measurement of Hubble constant. We assume a conservative eccentricity $e_0=0.2$ at 0.1 Hz. For BNS and NSBH, the eccentricity mainly helps for the distance inference. While for BBH, the source localization is improved significantly from eccentricity so that AEDGE can observe the golden dark BBH. We randomly select 5 and 10 golden dark sirens in the BNS, NSBH, and BBH catalogs (assuming $e_0=0.2$). We assume that the number of golden events that can be tracked by AEDGE is at least 1 per year. Thus 5 and 10 golden dark sirens correspond to 5 and 10 years data-taking period.

To measure the Hubble constant we need to assess the total distance errors of the golden dark sirens. We can safely neglect the weak lensing contribution since the golden dark sirens mainly reside in the low redshift region. However, the peculiar velocity is prominent at small $z$.
We use the fitting formula~\cite{Kocsis:2005vv},
\begin{equation}
\left(\frac{\Delta d_L(z)}{d_L(z)}\right)_{\rm pec}=\left[1+\frac{c(1+z)^2}{H(z)d_L(z)}\right]\frac{\sqrt{\langle v^2\rangle}}{c} \,,
\end{equation}
here we set the peculiar velocity value to be 500 km/s, in agreement with average values observed in galaxy catalogs. The final uncertainty of $d_L$ is the sum of the errors from GW parameter estimation and from the peculiar velocity in quadrature.
We assume the $\Lambda$CDM model with two free parameters $H_0,~\Omega_m$. However, at low redshift, $\Omega_m$ is poorly constrained. To get the posteriors of $H_0$, we run Markov-Chain Monte-Carlo (MCMC) by using the package {\sc Cobaya}~\cite{Torrado:2020dgo,2019ascl.soft10019T}. The marginalized statistics of the parameters and the plots are produced by the Python package {\sc GetDist}~\cite{Lewis:2019xzd}. 
	
\begin{figure}
\centering
\includegraphics[width=0.9\textwidth]{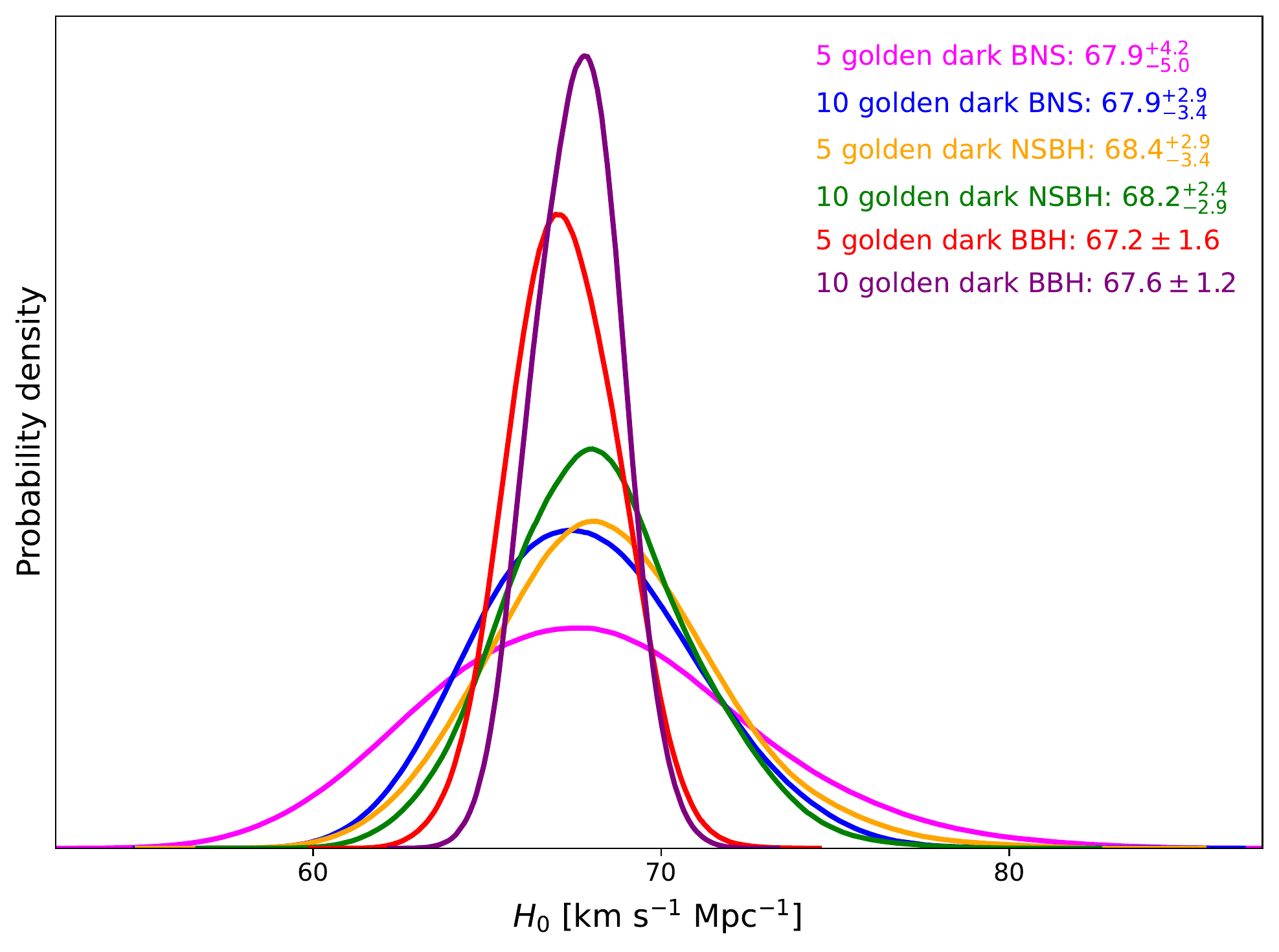}
\caption{The measurements of the Hubble constant from 5 and 10 golden dark sirens detected by AEDGE. The numbers are the mean values with 68\% limits of the errors.}
\label{fig:H0}
\end{figure}

Figure~\ref{fig:H0} shows the measurements of the Hubble constant from 5--10 golden dark BNS, NSBH, and BBH, which are randomly selected from the catalogs with $e_0=0.2$ in table~\ref{tab:np}. To avoid the bias introduced by the random selection, for each case, we repeat the process (the selection, random scattering, and MCMC) for 10 times. We choose the median one among the 10 repetition as the representative result. Our result suggests that with 5 (10) golden BNS, NSBH, BBH, AEDGE can constrain $H_0$ at 6.8\% (4.6\%), 4.6\% (3.9\%), and 2.4\% (1.8\%) precision levels, respectively. Note the total number of golden dark NSBH is around 40. If somehow AEDGE can track all of them during the observational time, the constraint of Hubble constant from NSBH can be further improved to $\sim 2\%$, according to $\sigma\sim 1/\sqrt{N}$ where $N$ is the number of the events. We find the golden dark BBH is more efficient than BNS and NSBH in constraining the Hubble constant. With 5--10 golden dark BBH one can obtain a 2 percent measurement of $H_0$ which is sufficient to arbitrate the Hubble tension. To obtain tens golden dark BBH, eccentricity is of great importance as shown in table~\ref{tab:np}. Our result shows eccentricity, which is more likely to exist in the mid-band, has great significance for the dark sirens with AEDGE as probes of the Universe.

\section{Conclusions and discussions \label{sec:conclusion}}

In this paper, we first investigate the eccentricity effects on the distance inference and source localization of the typical compact binaries observed by the atom interferometer AEDGE. We simulate 5 types of typical compact binaries in GWTC-3 with component mass ranging from $1-100~M_{\odot}$. We find that eccentricity can significantly improve the distance inference of all the typical binaries in the near face-on orientations in which case the distance largely degenerates with the inclination angle. The largest improvements correspond to 1.5--3 orders of magnitude for $e_0=0.4$. More importantly, eccentricity  
can improve greatly the localization of the typical BBH, most by a factor of 1.5--3 orders of magnitude. Generally, the heavier-mass binaries with higher eccentricity can achieve more improvements, which is consistent with the result in the LIGO/Virgo band~\cite{Sun:2015bva,Ma:2017bux,Pan:2019anf}. However, we find the improvements in the mid-band are much more significant than that in the LIGO/Virgo band. 

To predict the eccentricity effects on the GWs detected by AEDGE in the future, we simulate the catalogs of dark sirens which are within the detection range of AEDGE in 5 years, by assuming different eccentricities. Our results show that with no eccentricity, the numbers of BNS, NSBH, and BBH ($z<2$) that pass the detection threshold of AEDGE are the order of $\mathcal{O}(10^2)$, $\mathcal{O}(10^3)$, and $\mathcal{O}(10^5)$, respectively. With nonvanishing eccentricities the numbers are slightly reduced due to the shrinkage of the inspiral phase. However, eccentricity can improve the overall distance inference of the GWs in the catalogs. Especially, the improvement of localization for BBH makes it possible to detect the golden dark BBH whose unique host galaxy can be identified. Regardless of eccentricity, the total number of golden dark sirens in the catalogs is around 50--60. This also provides an update to the forecast in Paper II (with $e_0=0$), by adopting the latest inferred merger rates and some refinements in the calculation. We forecast the constrains of the Hubble constant from 5--10 golden dark sirens with AEDGE. We find BBH is more efficient to measure the Hubble constant than BNS and NSBH. With 5--10 golden dark BBH one can obtain a 2 percent measurement of $H_0$ which is sufficient to arbitrate the Hubble tension. Since eccentricity is very crucial to the detection of golden dark BBH, our results convey an important message to the community that eccentricity has great significance for the dark sirens as precise probes of the Universe. 

Though inferior to BBH when constraining the Hubble constant, golden BNS and NSBH are much more promising to be detected regardless of the eccentricity. On the one hand, if somehow 40 golden dark NSBH can be tracked by AEDGE, a $2\%$ percent Hubble constant measurement can also be guaranteed to arbitrate the Hubble tension. On the other hand, the golden dark BNS and NSBH can serve as an early warning for the follow-up observation of the EM counterparts. In addition, the host galaxies indicated by the EM counterparts can be compared with the one previously identified through the precise localization. This provides a validity check of the host galaxy identification of the golden dark sirens, which is informative to the golden dark BBH.

When applying the golden dark sirens in the catalogs to measuring the Hubble constant, we adopt the catalogs with a uniform eccentricity $e_0=0.2$. This is a conservative value we assume for the eccentricity (averagely) at 0.1 Hz. The exact distribution of eccentricity in the mid-band is still very uncertain. However, we found that the choice of eccentricity dose not influence the ability of detecting the golden dark BNS and NSBH. For BBH, except in the circular case, the detection of golden dark BBH is guaranteed. The different choice of the nonvanishing eccentricity (from $e_0=0.1$ to 0.4) dose not influence the detection of golden dark BBH seriously.

In this paper, when constructing the catalogs of GWs we adopt the median value of the merger rates which are inferred from GWTC-3. Taking account of the uncertainties of the merge rates could enlarge or reduce the population of GWs in the catalogs with AEGDE. As discussed in Paper II, since we can only track a very limited number of events in the resonant mode of AEDGE, the uncertainty of the population of GWs has small influence on our final results. This argument also holds true for the different assumptions of the number density of the galaxies. However, even assuming a very large $n_g= 0.1~\rm Mpc^{-3}$, we can still observe more than 5 golden dark sirens for each type of binaries. In addition, as discussed in Paper II, the clustering and grouping of galaxies make it much easier to infer the redshift of GWs from the cluster and group of the host galaxy instead of the host galaxy itself~\cite{Yu:2020vyy}. This means that our estimation is somehow very conservative. 

From table~\ref{tab:np} AEDGE can observe hundreds of dark sirens with less than 10 potential host galaxies within the localized region. These dark sirens are also very useful by weighting the probability of hosting the GWs among these few potential galaxies.  Though the constraints on cosmological parameters from one such event could be looser than that from a golden dark siren, a large number of these events combined may still provide comparable or better measurements of the Hubble constant. Moreover, with these evens at higher redshift we can measure not only the local Hubble constant but also the matter density parameter and equation of state of dark energy, etc.

 
\appendix
\section{Derivation of AEDGE antenna pattern functions \label{app:F}}

The GW strain tensor can be decomposed in terms of 
\begin{equation}
h_{ij}(t)=h_+(t)\mathbf{e}_{ij}^++h_\times(t)\mathbf{e}_{ij}^{\times}\,,
\label{eq:hij}
\end{equation}
here $\mathbf{e}_{ij}^{+,\times}$ are  the polarization tensors with $\mathbf{e}_{ij}^+=\mathbf{u}_i\mathbf{u}_j-\mathbf{v}_i\mathbf{v}_j$, and $\mathbf{e}_{ij}^\times=\mathbf{u}_i\mathbf{v}_j+\mathbf{v}_i\mathbf{u}_j$. 
We first assume the polarization angle $\psi=0$. For the source located at $(\theta,\phi)$ in the heliocentric ecliptic coordinate system, the bases of GW polarization tensors are
\begin{align}
\mathbf{u}=&(\cos\theta\cos\phi, \cos\theta\sin\phi, -\sin\theta) \,,\\
\mathbf{v}=&(\sin\phi, -\cos\phi, 0) \,.
\label{eq:uv}
\end{align}
For the single-baseline detector AEDGE, the detector response tensor $D_{ij}$ from the baseline direction unit vector $\mathbf{a}(t)$ is
\begin{equation}
D_{ij}=\frac{1}{2}\mathbf{a}_i(t)\mathbf{a}_j(t) \,.
\label{eq:Dij}
\end{equation}
We parameterize the detector location on the orbit around the earth by a unit vector in the geocentric coordinates,
\begin{equation}
\mathbf{r_{AI}}_{,0}(t)=(\cos\phi_a(t),\sin\phi_a(t),0) \,,
\end{equation}
where $\phi_a(t)=2\pi t /T_{\rm AI}+\phi_0$ is the azimuthal orbit angle around the Earth. $T_{\rm AI}=10$ days is the orbit period of AEDGE around the Earth. Then the baseline direction of AEDGE in geocentric coordinates is 
\begin{equation}
\mathbf{a}_0(t)=(-\sin\phi_a(t),\cos\phi_a(t),0) \,.
\end{equation}
We need to transform $\mathbf{a}_0(t)$ in the geocentric coordinates to $\mathbf{a}(t)$ in the heliocentric coordinates,
\begin{align}
\mathbf{a}(t)=
\begin{pmatrix}
\cos\phi_{\rm Ea}(t) & -\sin\phi_{\rm Ea}(t) & 0 \\
\sin\phi_{\rm Ea}(t) & \cos\phi_{\rm Ea}(t) & 0 \\
0 & 0 & 1
\end{pmatrix} \cdot
\begin{pmatrix}
\cos\theta_{\rm inc} & 0 & -\sin\theta_{\rm inc} \\
0 & 1 & 0 \\
\sin\theta_{\rm inc} & 0 & \cos\theta_{\rm inc}
\end{pmatrix}
\cdot
\mathbf{a}_0(t) \,.
\end{align}
The azimuthal angle of the Earth’s orbit around the Sun is $\phi_{\rm Ea}(t)=2\pi t/(1 \rm yr)+\phi_0'$. The inclination $\theta_{\rm inc}=28.5^{\circ}$ is the angle between the orbit plane of AEDGE around the Earth and the ecliptic.
Then the observed waveform is given by
\begin{equation}
h(t)\equiv D_{ij}h_{ij}=h_+(t)F_+(t)+h_\times(t)F_\times(t) \,.
\end{equation}
Since we assume $\psi=0$ above, we get $F_+(t,\psi=0)=D_{ij}(t)\mathbf{e}_{ij}^+$ and $F_\times(t,\psi=0)=D_{ij}(t)\mathbf{e}_{ij}^\times$. Then taking the polarization angle into account we get
\begin{align}
F_+(t)=&\cos(2\psi)F_+(t,\psi=0)-\sin(2\psi)F_\times(t,\psi=0) \,, \\
F_\times(t)=&\sin(2\psi)F_+(t,\psi=0)+\cos(2\psi)F_\times(t,\psi=0) \,.
\end{align}
These are the antenna pattern functions of AEDGE.

In the frequency domain, we need to derive the relation between time and GW frequency, $t(f)$. We numerically solve the orbital evolution equation in~\cite{Yunes:2009yz} for the eccentric cases. We use the frequency of the dominant quadrupole mode $\ell=2$ as the variable in $t(f)$. Then the antenna pattern functions for $\ell=2$ are $F_{+,\times}(t(f))$. Since different modes enter frequency band at different time, the time at the frequency of nonquadrupole modes should be related to that at quadrupole mode by $t(f)=t(2f_\ell /\ell)$. Therefore, the antenna pattern functions 
for $\ell$ mode should be $F_{+,\times}(t(2f/\ell))$. The total strain is $h=\sum_\ell h_\ell=\sum_\ell h_{\ell+}F_+(t(2f/\ell))+h_{\ell\times}F_\times(t(2f/\ell))$.

\section{Supplementary results for the typical events\label{app:sup}}
\begin{figure}[h]
\centering
	\begin{subfigure}[b]{0.3\textwidth}
	\centering
	\includegraphics[width=\textwidth]{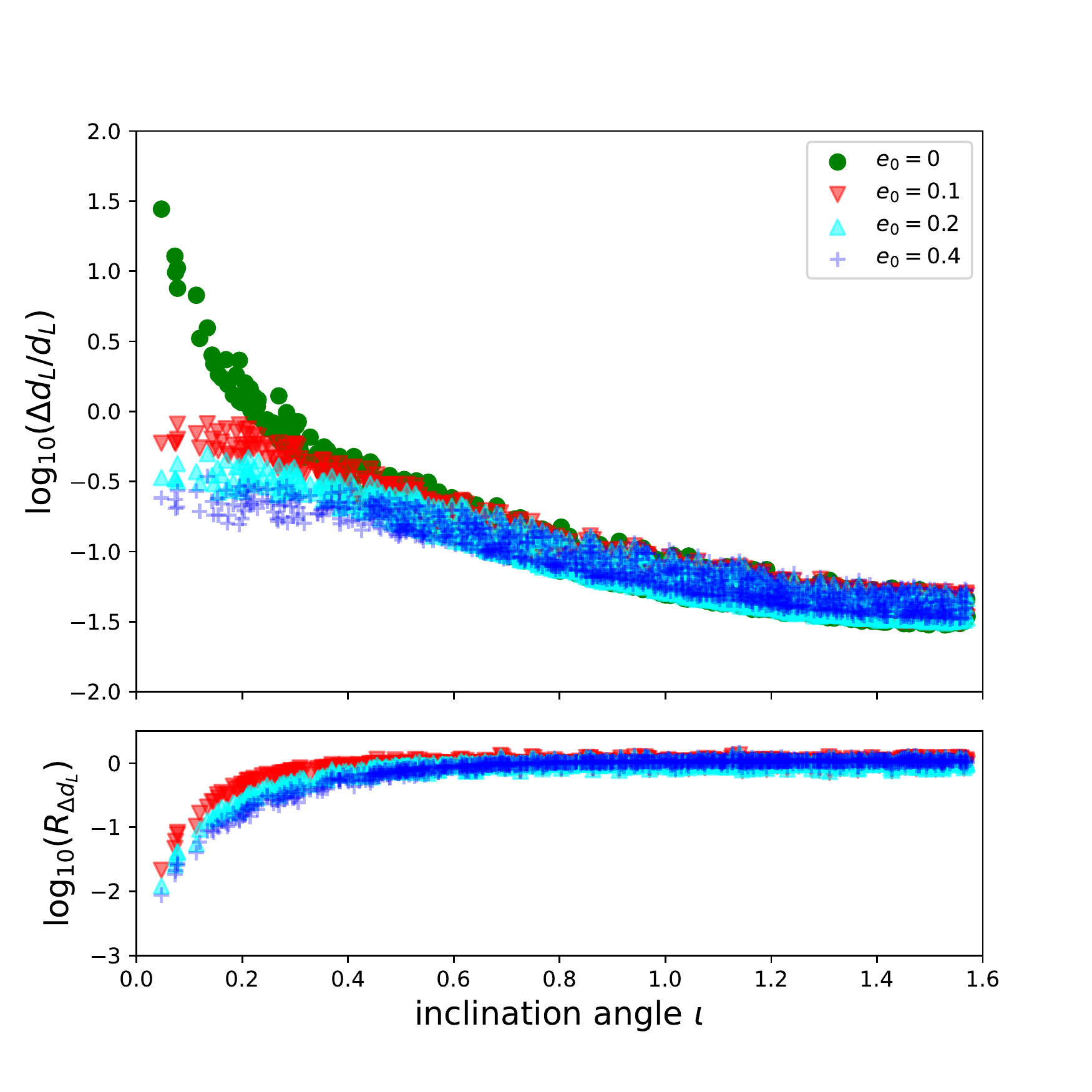}
	\caption{GW170817-like BNS}
	\end{subfigure}
	\begin{subfigure}[b]{0.3\textwidth}
	\centering
	\includegraphics[width=\textwidth]{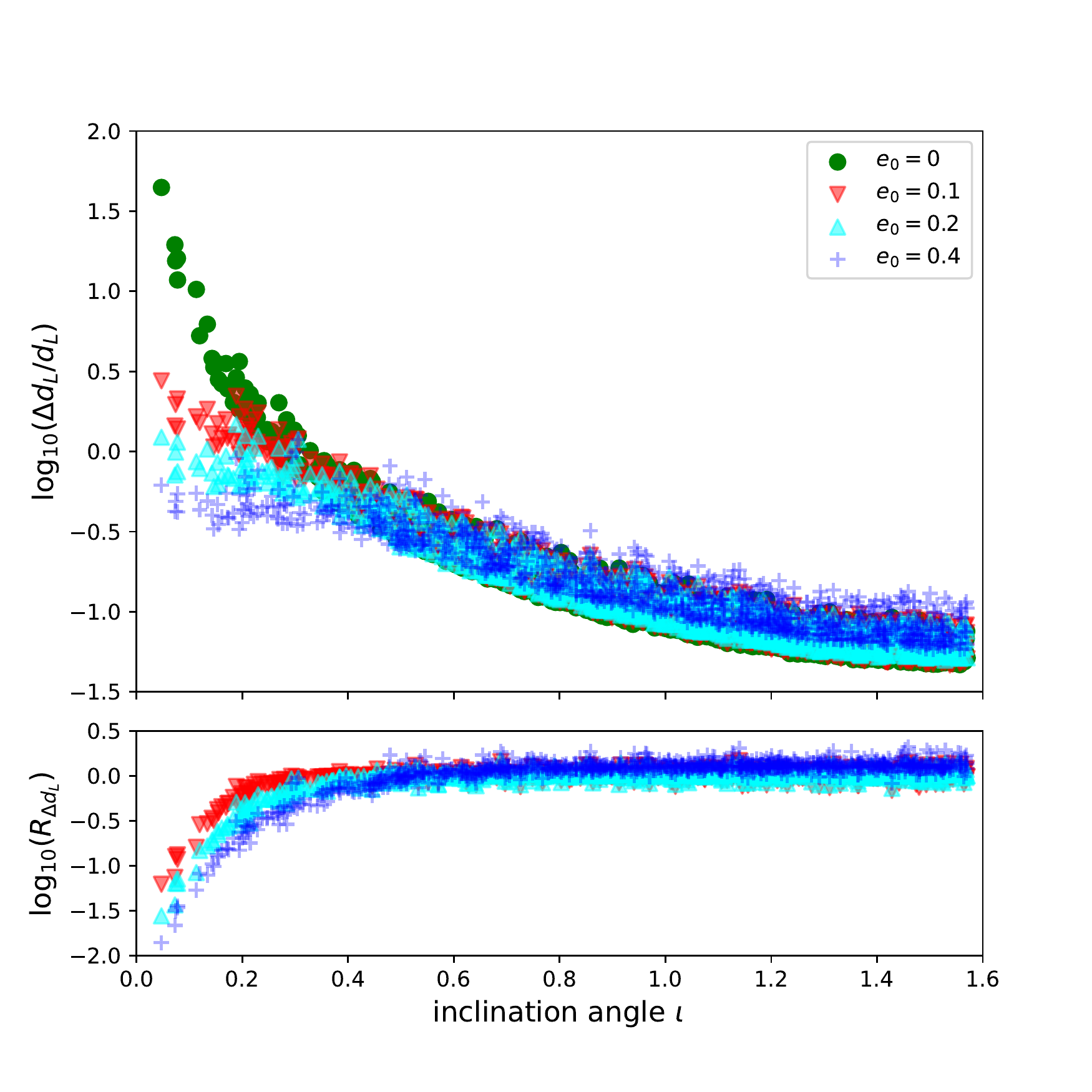}
	\caption{GW200105-like NSBH}
	\end{subfigure}
	\begin{subfigure}[b]{0.3\textwidth}
	\centering
	\includegraphics[width=\textwidth]{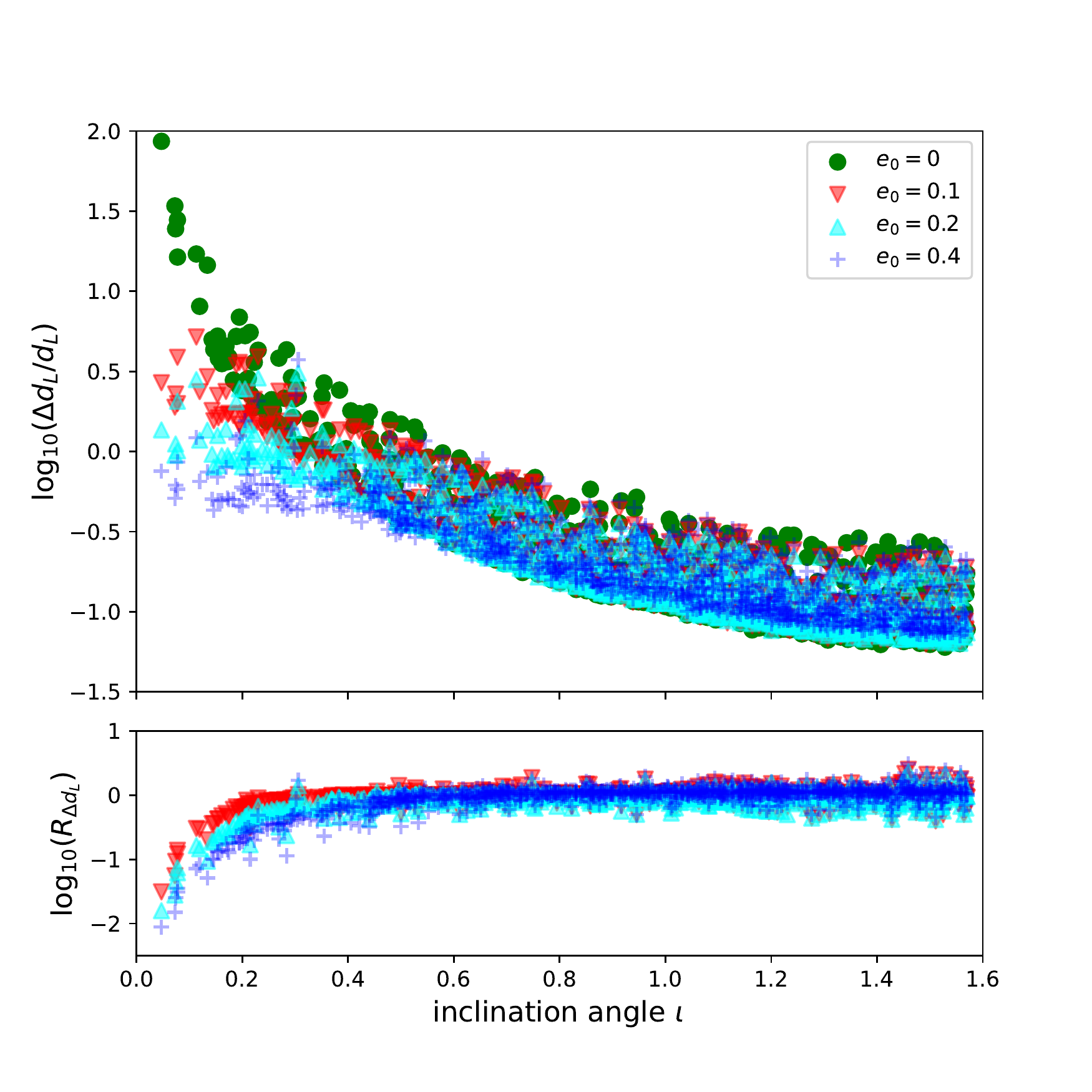}
	\caption{GW191129-like light BBH}
	\end{subfigure}
	\begin{subfigure}[b]{0.3\textwidth}
	\centering
	\includegraphics[width=\textwidth]{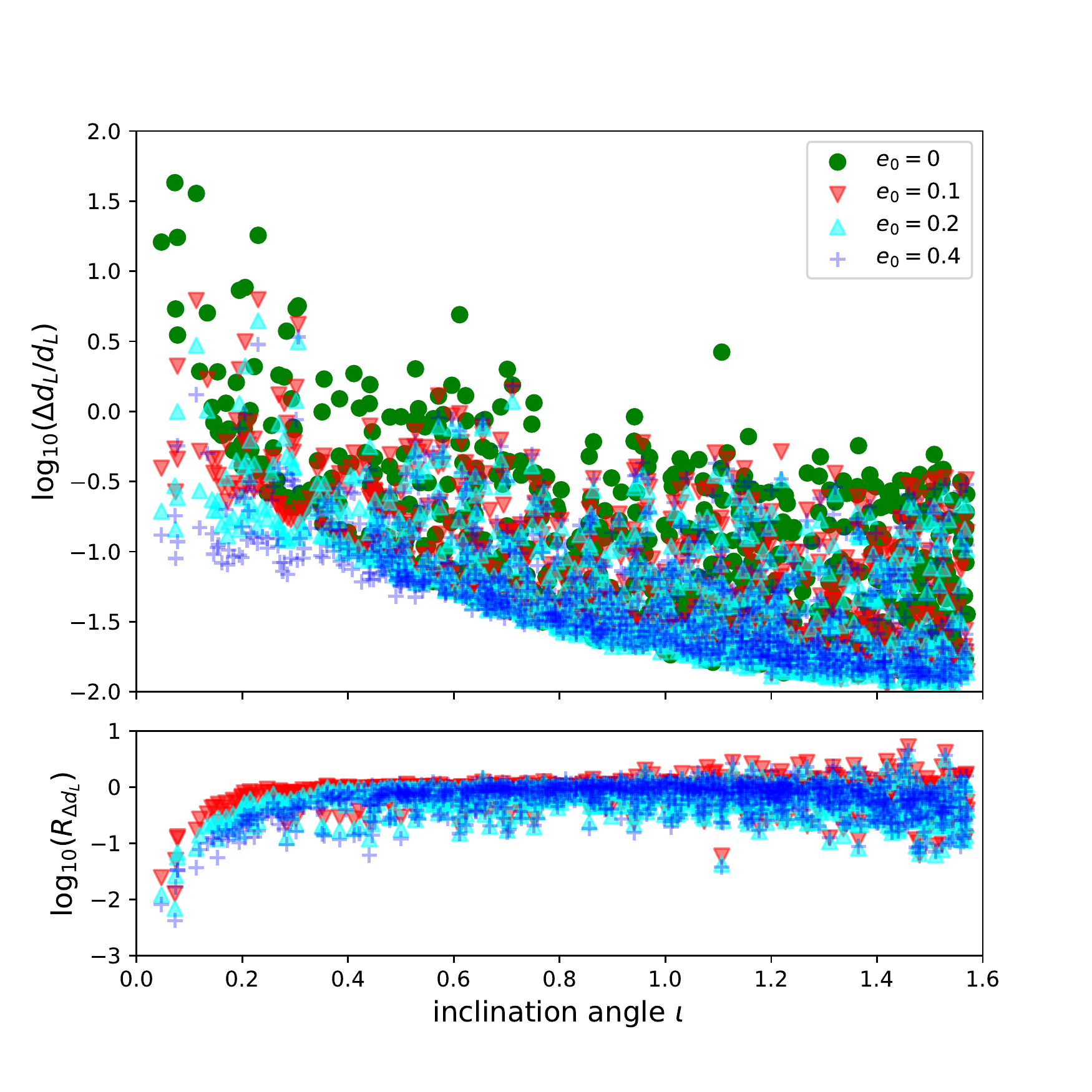}
	\caption{GW150914-like medium BBH}
	\end{subfigure}
	\begin{subfigure}[b]{0.3\textwidth}
	\centering
	\includegraphics[width=\textwidth]{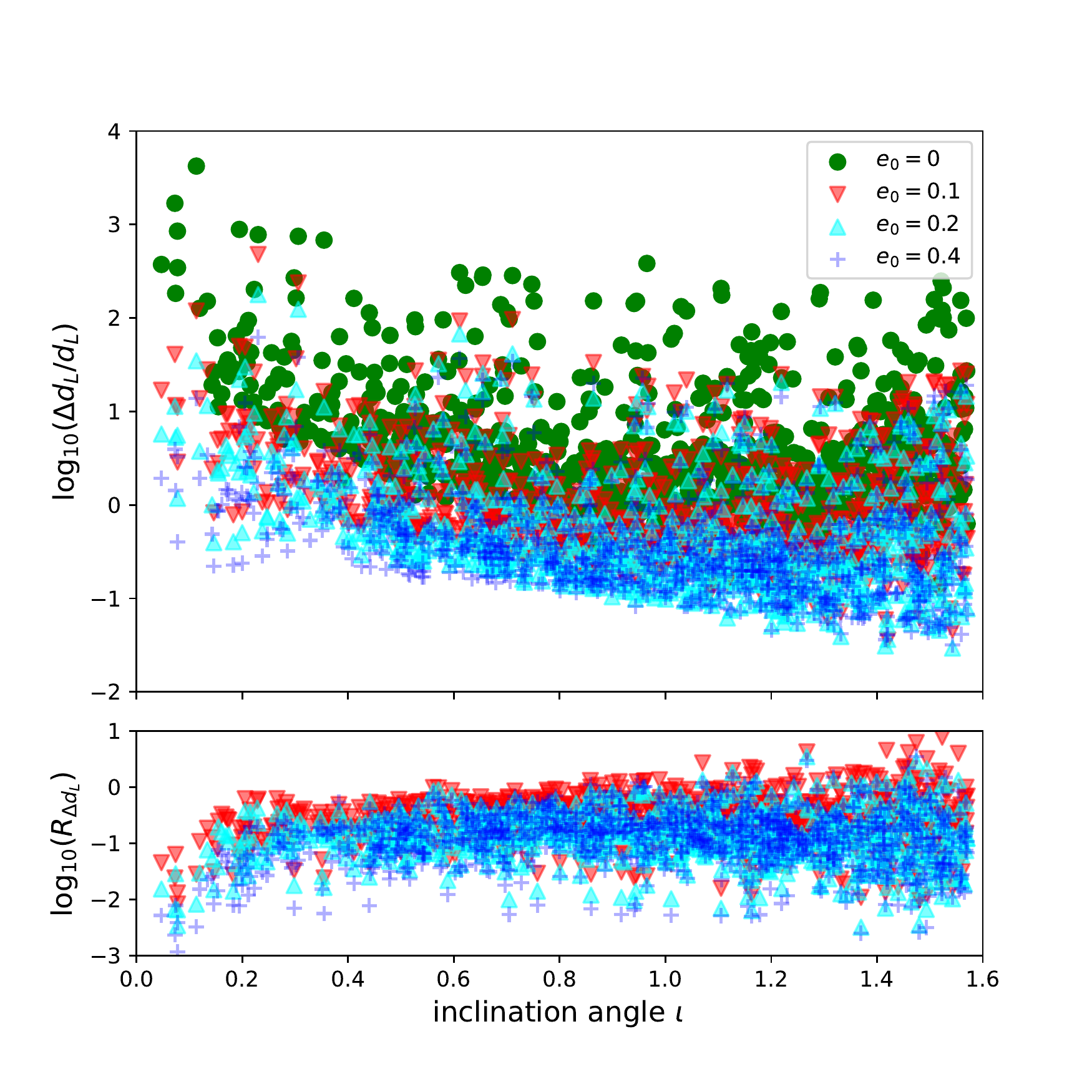}
	\caption{GW190426-like heavy BBH}
	\end{subfigure}
\caption{The distance inference of the five typical binaries with various eccentricities.}
\label{fig:err_dL_sum}
\end{figure}

\begin{figure}[h]
\centering
	\begin{subfigure}[b]{0.3\textwidth}
	\centering
	\includegraphics[width=\textwidth]{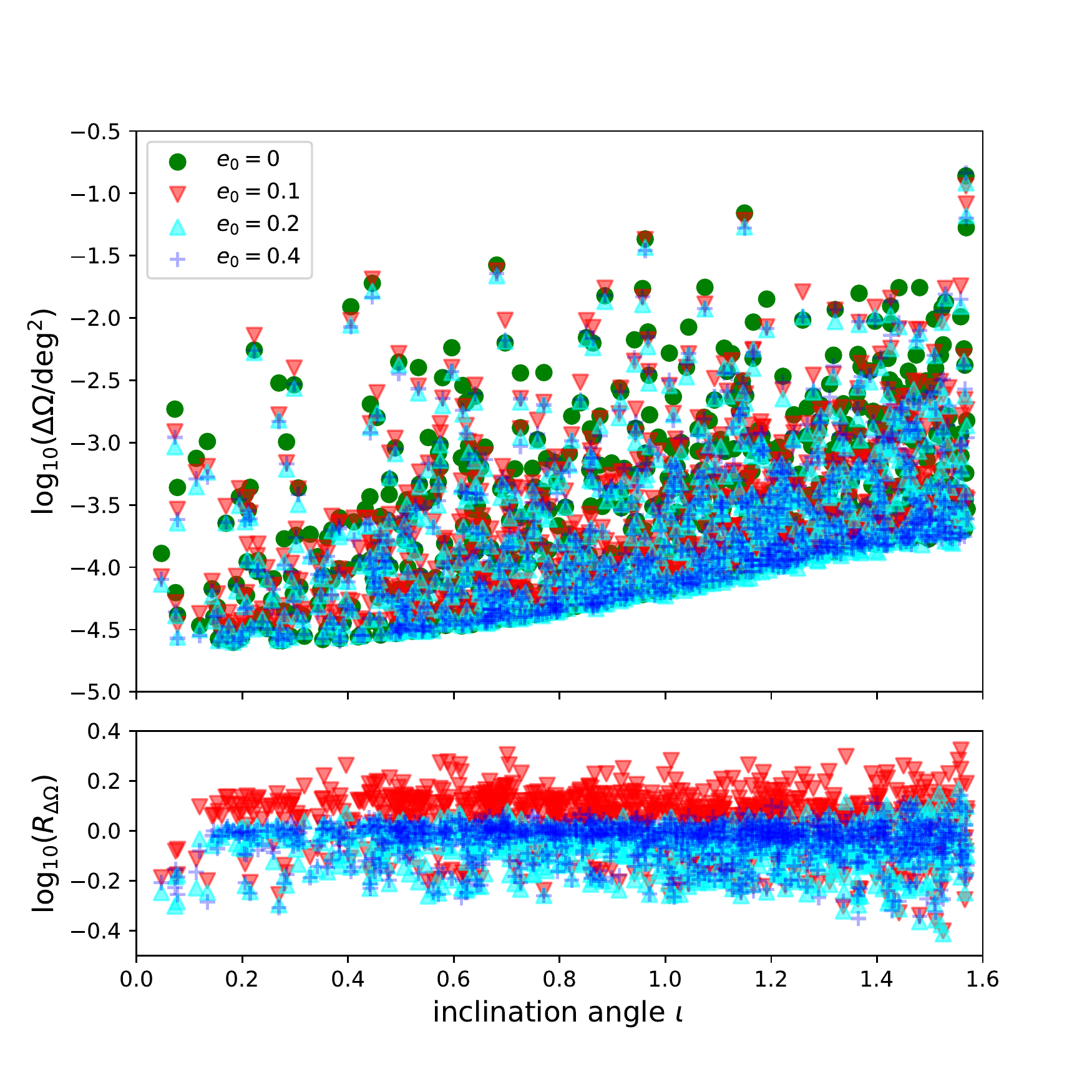}
	\caption{GW170817-like BNS}
	\end{subfigure}
	\begin{subfigure}[b]{0.3\textwidth}
	\centering
	\includegraphics[width=\textwidth]{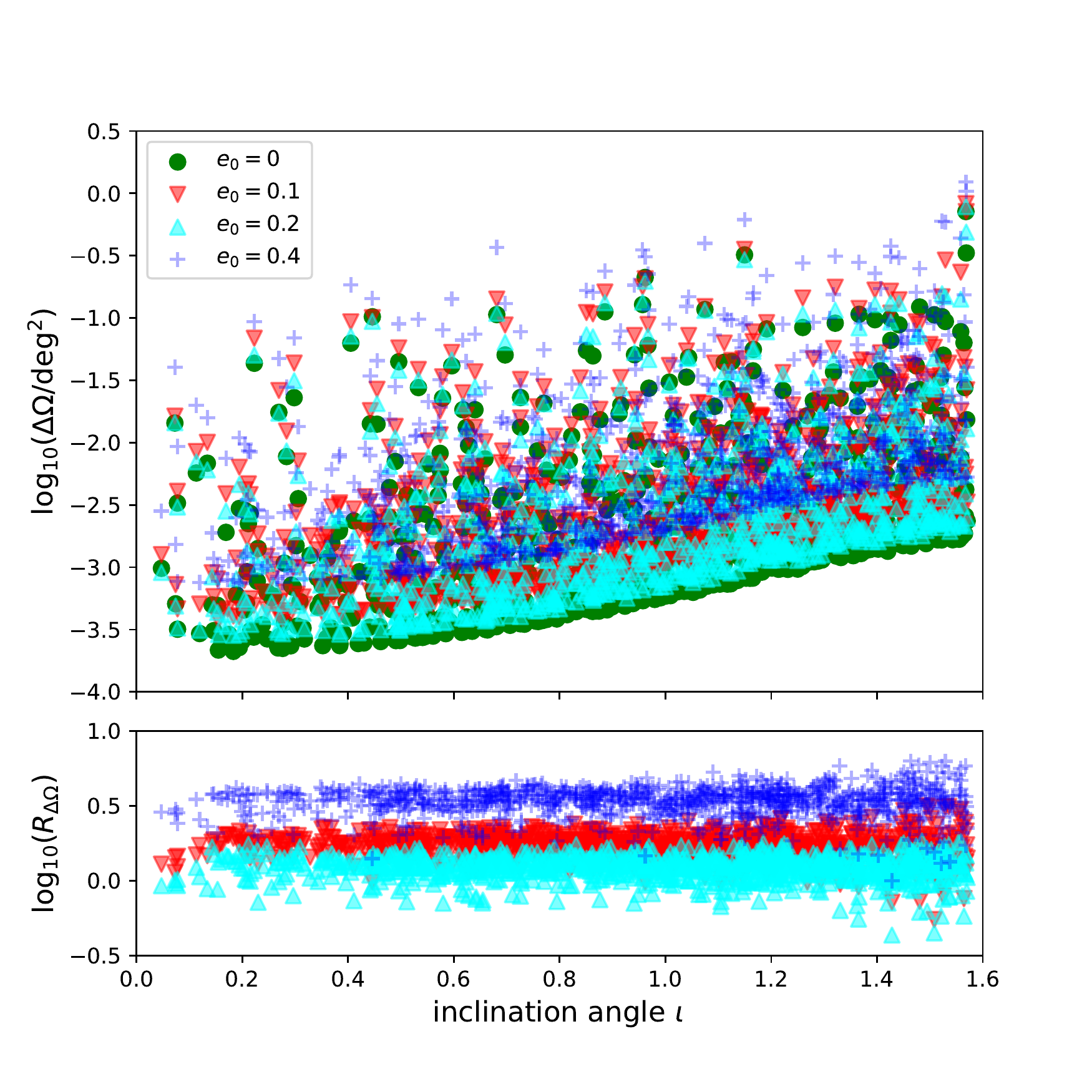}
	\caption{GW200105-like NSBH}
	\end{subfigure}
	\begin{subfigure}[b]{0.3\textwidth}
	\centering
	\includegraphics[width=\textwidth]{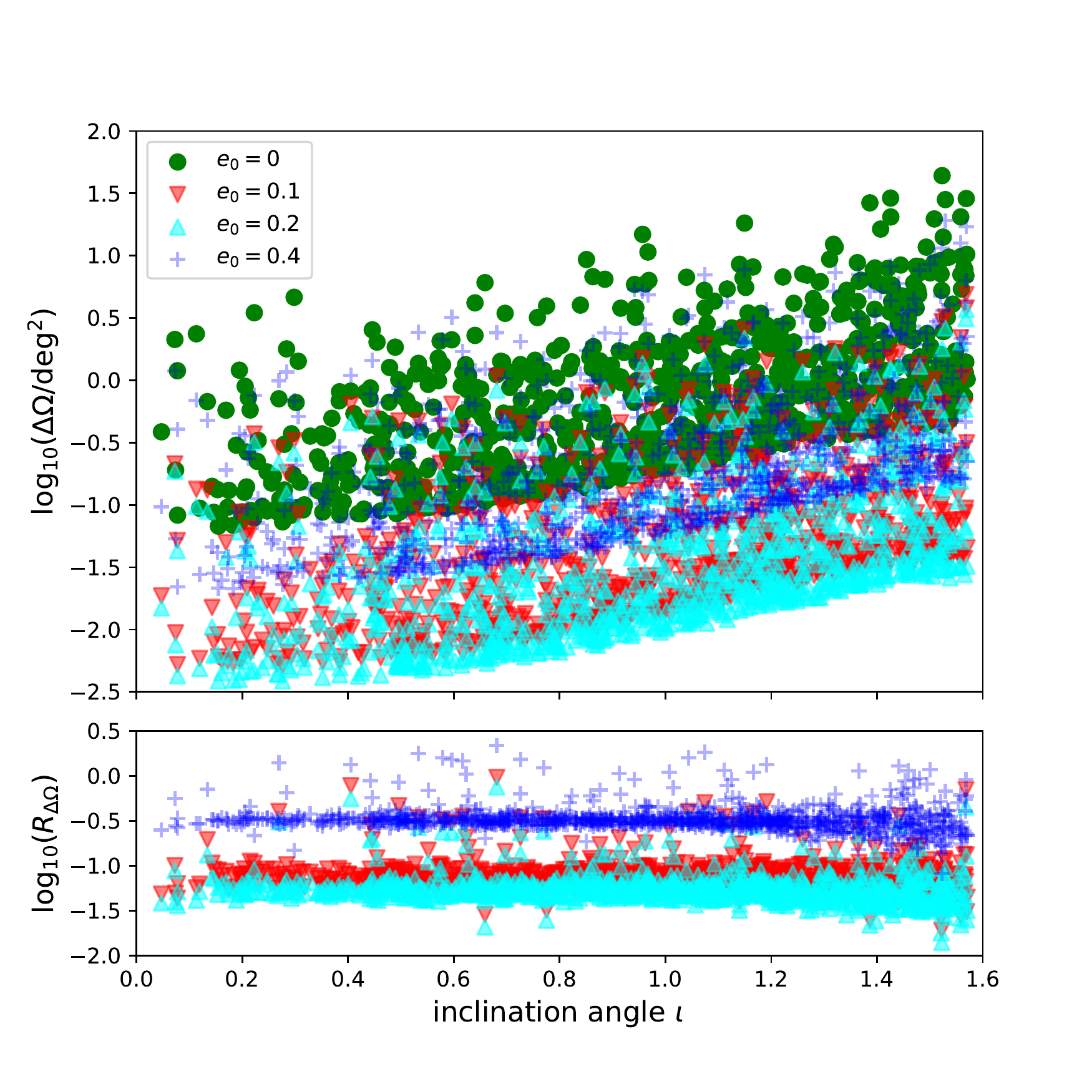}
	\caption{GW191129-like light BBH}
	\end{subfigure}
	\begin{subfigure}[b]{0.3\textwidth}
	\centering
	\includegraphics[width=\textwidth]{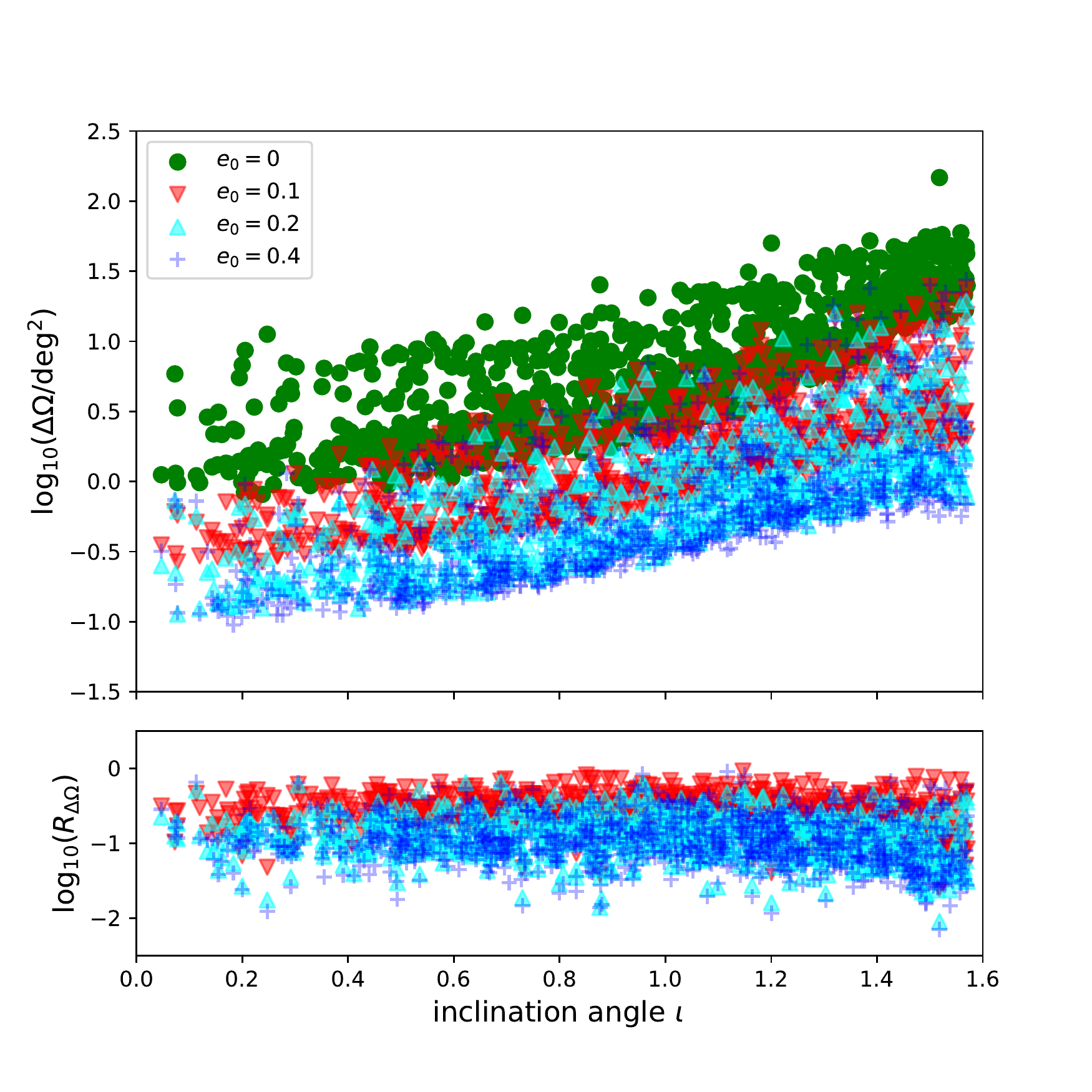}
	\caption{GW150914-like medium BBH}
	\end{subfigure}
	\begin{subfigure}[b]{0.3\textwidth}
	\centering
	\includegraphics[width=\textwidth]{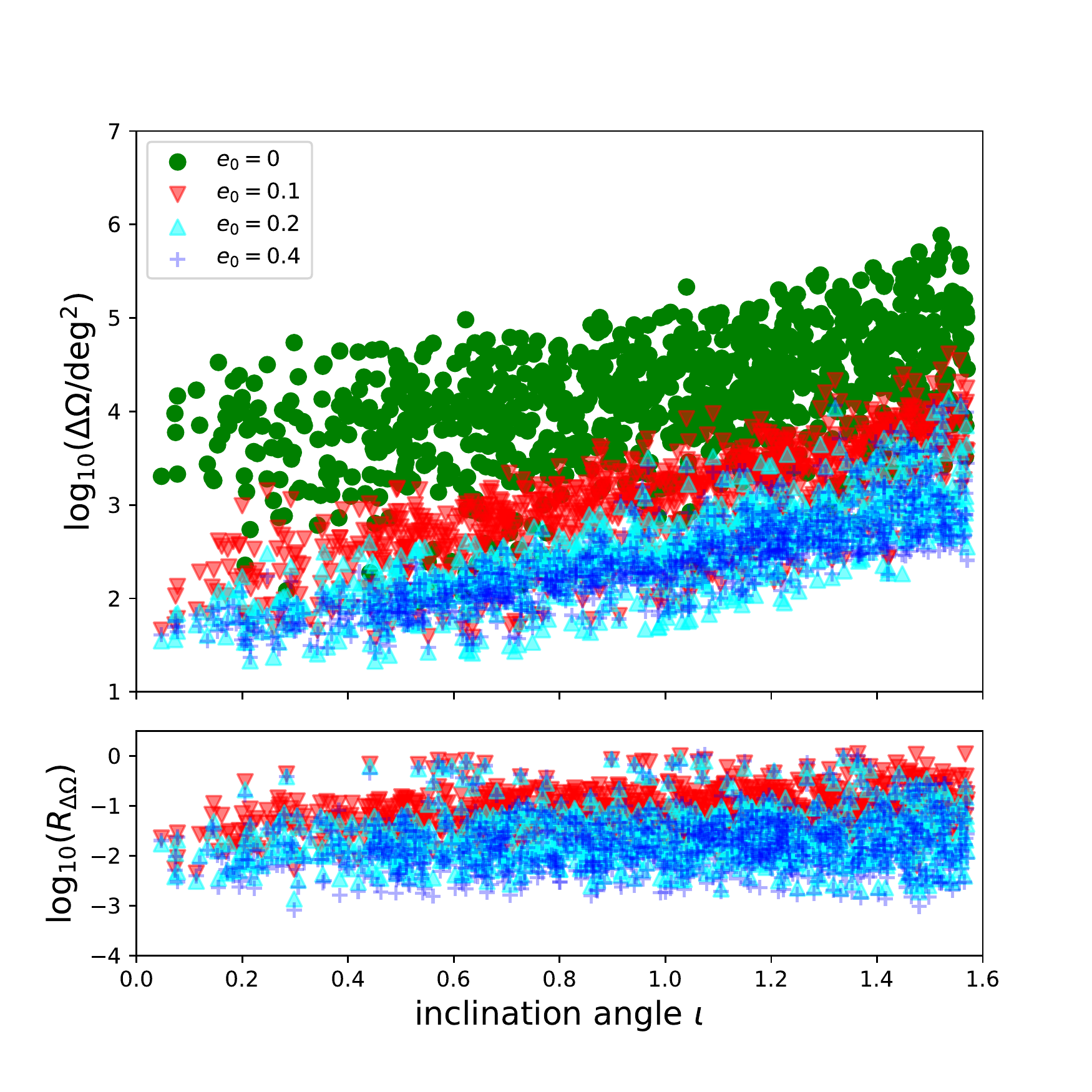}
	\caption{GW190426-like heavy BBH}
	\end{subfigure}
\caption{The source localization of the five typical binaries with various eccentricities.}
\label{fig:err_Omega_sum}
\end{figure}

\begin{figure}[h]
\centering
	\begin{subfigure}[b]{0.32\textwidth}
	\centering
	\includegraphics[width=\textwidth]{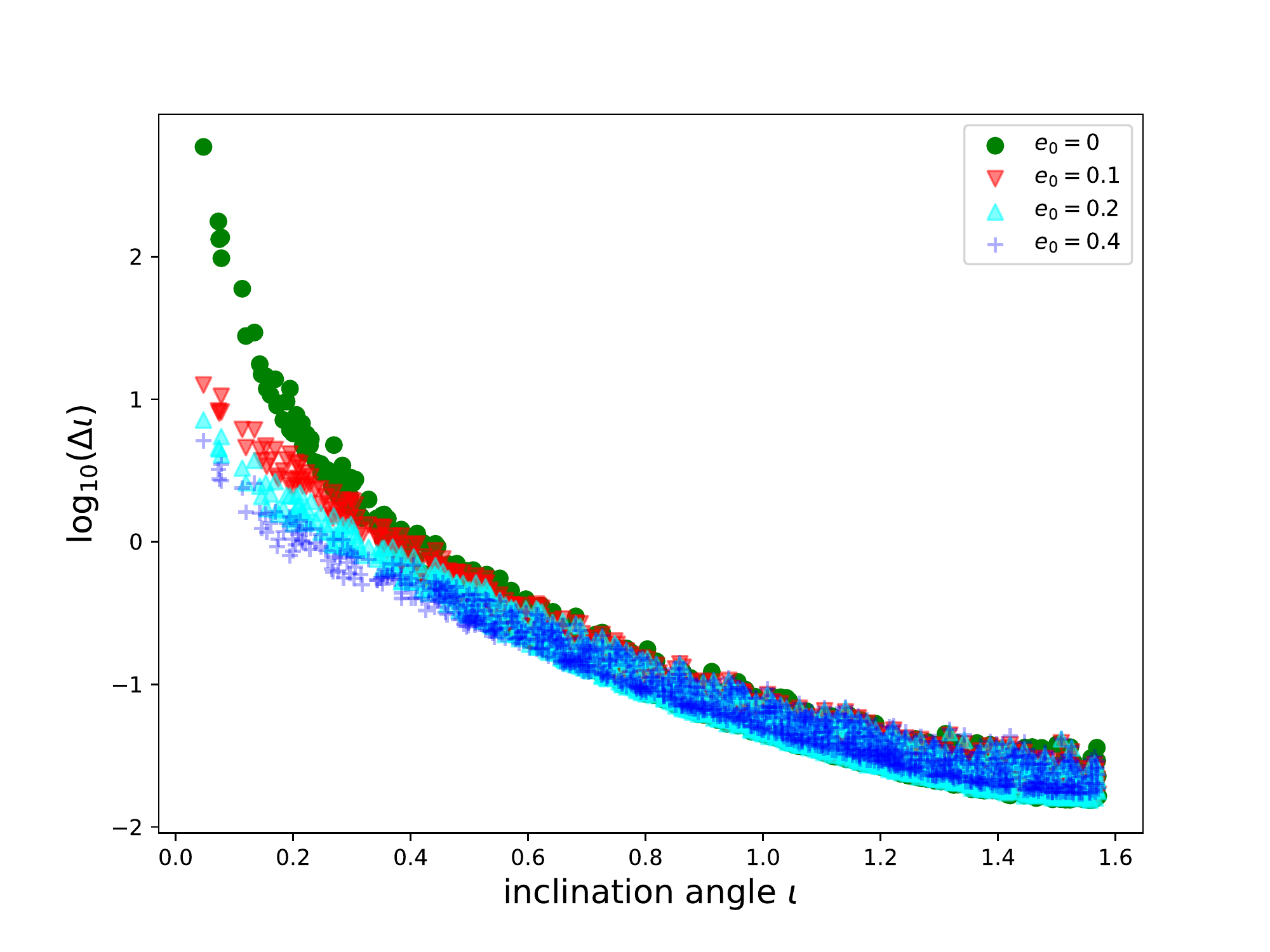}
	\caption{The error of inclination for GW170817-like BNS}
	\label{fig:diota}
	\end{subfigure}
	\begin{subfigure}[b]{0.32\textwidth}
	\centering
	\includegraphics[width=\textwidth]{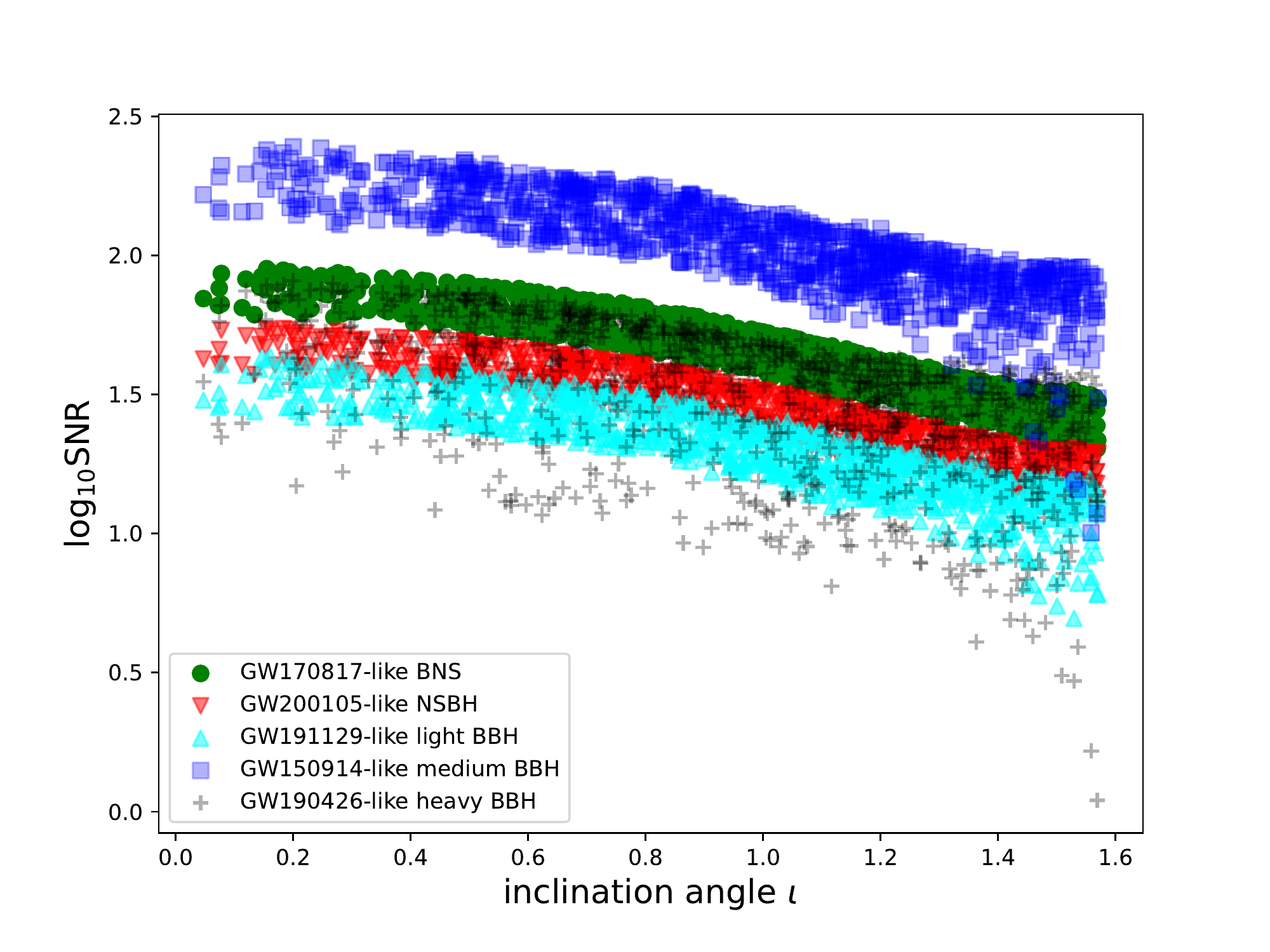}
	\caption{The SNR of the typical events with $e_0=0.2$}
	\label{fig:SNR_e02}
	\end{subfigure}	
	\begin{subfigure}[b]{0.32\textwidth}
	\centering
	\includegraphics[width=\textwidth]{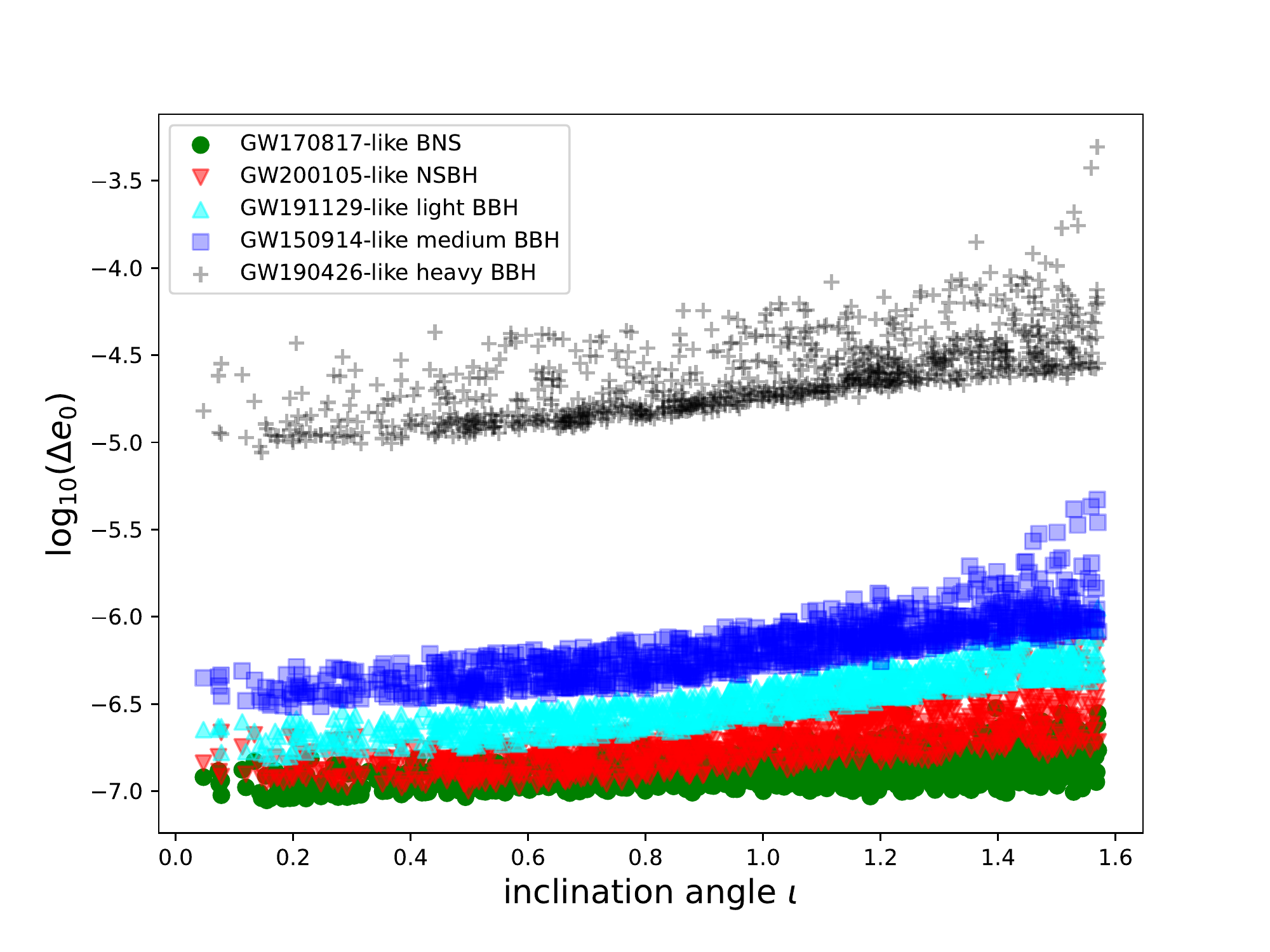}
	\caption{The error of $e_0$ of the typical events with $e_0=0.2$}
	\label{fig:erre0_e02}
	\end{subfigure}	
\caption{Some supplementary results.}
\label{fig:sup}
\end{figure}

Here we summarize some additional results as supplementary to section~\ref{sec:typical}. For the eccentric cases, we only show the results with $e_0=0, 0.1, 0.2$ and 0.4. (1) We first show the distance inference and source localization of all the five typical binaries in figure~\ref{fig:err_dL_sum} and~\ref{fig:err_Omega_sum}. (2) We have pointed out in the main text that eccentricity can largely break the degeneracy between distance $d_L$ and inclination $\iota$ in the near face-on orientations. We expect the errors of inclination angle can also be significantly reduced in small $\iota$. As shown in figure~\ref{fig:diota}, we can see the clear significant improvements of $\iota$ in the near face-on orientations. (3) To compare the SNR between these typical binaries, we choose $e_0=0.2$ as representative shown in figure~\ref{fig:SNR_e02}. As expected, with larger inclination the SNR is lower. In the main text, we concluded that a heavier mass binary can achieve more improvements from eccentricity. We can see this is not relevant to the SNR of the typical binaries. The heavy BBH gets the largest improvement for distance inference and localization. But its SNR is not the highest. (4) Though in this paper 
we do not show the error of eccentricity, we are still curious about how precisely the eccentricity can be constrained by AEDGE in the mid-band. As shown in figure~\ref{fig:erre0_e02}, we plot the error of $e_0$ against inclination angle for these five typical binaries by assuming $e_0=0.2$ as a representative. We find the eccentricity can be constrained very precisely with $\Delta e_0\sim 10^{-5}-10^{-7}$ for these binaries.

\acknowledgments
This work is supported by National Research Foundation of Korea 2021R1A2C2012473  and 2021M3F7A1082056.
RGC is supported by National Key Research and Development Program of China Grant No. 2020YFC2201502, the National Natural Science Foundation of China Grants No.11821505, No.11991052, No.11947302 and by the Strategic Priority Research Program of the Chinese Academy of Sciences Grant No.XDB23030100, and the Key Research Program of FrontierSciences of CAS.

\bibliographystyle{JHEP}
\bibliography{ref}

\end{document}